\documentclass[usenatbib]{mn2e}
\usepackage{amsfonts}
\usepackage{txfonts}
\usepackage{mathrsfs}
\usepackage{amssymb}
\usepackage{psfig}
\usepackage{graphicx,times}
\usepackage{natbib}
\usepackage{lscape}
\usepackage{textcomp}
\usepackage{graphicx}
\usepackage{epsfig}
\usepackage{color}

\footnotesize

\title{Chemical Evolution of HC$_3$N in Dense Molecular Clouds}

\author[Nai-Ping Yu, Jun-Jie Wang and Jin-Long Xu]{Naiping Yu$^{1,2}$\thanks{E-mail: npyu@bao.ac.cn},
Jun-Jie Wang$^{1,2}$ ,Jin-Long Xu$^{1}$ \\
$^{1}$National Astronomical Observatories, Chinese Academy of
Sciences, Beijing 100012, China\\
$^{2}$NAOC-TU Joint Center for Astrophysics, Lhasa 850000, China}

\begin{document}
\maketitle \pagestyle{plain}

\begin{abstract}
We investigated the chemical evolution of HC$_3$N in six dense molecular clouds, using archival available data from the Herschel infrared Galactic Plane Survey (Hi-GAL) and the Millimeter Astronomy Legacy Team Survey at 90 GHz (MALT90). Radio sky surveys of the Multi-Array Galactic Plane Imaging Survey (MAGPIS) and the Sydney University Molonglo Sky Survey (SUMSS) indicate these dense molecular clouds are associated with ultracompact HII (UCHII) regions and/or classical HII regions. We find that in dense molecular clouds associated with normal classical HII regions, the abundance of HC$_3$N begins to decrease or reaches a plateau when the dust temperature gets hot. This implies UV photons could destroy the molecule of HC$_3$N. On the other hand, in the other dense molecular clouds associated with UCHII regions, we find the abundance of HC$_3$N increases with dust temperature monotonously, implying HC$_3$N prefers to be formed in warm gas. We also find that the spectra of HC$_3$N (10-9) in G12.804-0.199 and RCW 97 show wing emissions, and the abundance of HC$_3$N in these two regions increases with its nonthermal velocity width, indicating HC$_3$N might be a shock origin species. We further investigated the evolutionary trend of $N$(N$_2$H$^+$)/$N$(HC$_3$N) column density ratio, and found this ratio could be used as a chemical evolutionary indicator of cloud evolution after the massive star formation is started.
\end{abstract}



\section{Introduction}
Carbon-chain species account for a substantial fraction of the interstellar molecules observed so far. They are prone to be depleted onto dust grains when the gas is cold, and destroyed by UV radiations (Sakai \& Yamamoto 2013). In star-forming regions, many carbon-chain species could be used as ``chemical clocks" to trace star formation (e.g. Suzuki et al. 1992; Hirota et al. 2009). Cyanopolyynes (HC$_{2n+1}$N) are one of the representative carbon-chain species. Since the first detection of interstellar cyanoacetylene (HC$_3$N) by Turner (1971) in Sgr B2, cyanopolyynes have been found to be ubiquitously in our Galactic interstellar medium (ISM) (e.g., Cernicharo \& Gu$\acute{e}$lin 1996; Takano et al.
1998; Crovisier et al. 2004). Previously, long chain cyanopolyynes were believed to be abundant in cold dark clouds. In hot cores, they could not be formed efficiently (Millar 1997). However, long chain cyanopolyynes of HC$_5$N, HC$_7$N and HC$_9$N were detected by Sakai et al. (2008) in the protostar IRAS 04368+2557L1527. They proposed a new chemistry called ``warm carbon-chain chemistry (WCCC)" in a warm and dense region near the low-mass protostars. Hassell et al. (2008) then made a chemical model of this region. Their calculations show the cyanopolyynes abundance enrichment in the gas phase as the grains warm up to 30 K. Chapman et al. (2009) further presented that cyanopolyynes could be formed under a hot core condition and show as ``chemical clocks" to determine the age of hot cores.

HC$_3$N is the simplest form of cyanopolyynes. This molecule traces both dense and warm gas. In warm gas, it could be formed from CH$_4$ (Hassel et al. 2008) and/or C$_2$H$_2$ (Chapman et al. 2009) evaporated from grain mantles. Sanhueza et al. (2012) found the median value of HC$_3$N column density increases as a function of clump evolutionary stage. Taniguchi et al. (2016) observed three $^{13}$C isotopologues of HC$_3$N in L1527 and G28.28-0.36. They found the abundance of H$^{13}$CCCN and HC$^{13}$CCN are comparable, while HCC$^{13}$CN is more abundant. This result could be explained by that HC$_3$N might be formed from the neutral-neutral reaction between C$_2$H$_2$ and CN: C$_2$H$_2$ + CN $\longrightarrow$ HC$_3$N + H. The abundance of HC$_3$N can also be enhanced after the passage of shocks. Mendoza et al. (2018) found the abundance of HC$_3$N increases by a factor of 30 in the shocked region of L1157. Taniguchi et al. (2018) carried out observations of HC$_3$N and HC$_5$N toward 52 high-mass star-forming regions with the Nobeyama 45 m telescope. They found the spectra of some HC$_3$N show wing emissions, suggesting HC$_3$N is an outflow shock origin species.

The destruction of HC$_3$N could be caused by UV radiations. Yu \& Xu (2016) found the fractional abundances of HC$_3$N decrease as a function of Lyman continuum fluxes in a number of Red MSX (Midcourse Space Experiment) Sources (RMSs), indicating this molecule could be destroyed by UV photons. Urquhart et al. (2019) conducted a 3-mm molecular line survey towards 570 ATLASGAL (APEX Telescope Large Area Survey of the Galaxy) clumps. They found the detection rate of HC$_3$N (10-9) increases from the ``Quiescent" stage to the ``Protostellar" stage, and reaches a plateau in the ``Young Stellar Object (YSOs)" and ``HII region" stages. They guess that in the late two stages the formation of HC$_3$N is in equilibrium with its destruction by UV photons or other chemical reactions. Even to today, there are few researches about chemical evolution of HC$_3$N in massive star-forming regions. Previous studies mentioned above are surveys of massive clumps/cores in different giant molecular clouds (GMCs). The distances and initial conditions may be quite different
in different GMCs. For example, the measured $^{12}$C/$^{13}$C ratio ranges from $\sim$ 20 to $\sim$ 70, depending on the distance to the Galactic center (e.g. Savage et al. 2002). This might complicate quantitative comparisons and make statistical results not significant. In our previous paper (Yu et al. 2018 hereafter YXW18), we studied the chemical evolution
of N$_2$H$^+$ using data from MALT90 and Hi-Gal in six massive star-forming regions. Here we present our study of HC$_3$N instead of N$_2$H$^+$. The study of other molecules such as C$_2$H, HCO$^+$ and HNC will come in another paper. The distances and initial conditions of clumps could be regarded as the same in the same cloud, and thus a comparison of the chemical evolution in different clumps along the evolutionary sequence is valid. We introduce our data in Section 2, results and discussions are in Section 3, and finally we summarize in Section 4.

\section{Data and Analysis}
Our molecular line data of HC$_3$N (10-9) comes from MALT90, and the dust infrared data is from Hi-GAL. As the method described by YXW18, we first calculate the H$_2$ column density and dust temperature maps of these regions, and then the column density and abundance maps of HC$_3$N. Here we give a brief introduction.

The Hi-GAL data set is comprised of 5 continuum images of the Milky Way Galaxy using the PACS (70 and 160 $\mu$m) and SPIRE (250, 350 and 500 $\mu$m) instruments. Following the steps described by Wang et al. (2015), we made H$_2$ column density and dust temperature maps of each region through the method of spectral energy distribution (SED). After removing the background and foreground emissions, we convolved all the images of Hi-GAL to a spatial resolution of 45$^\prime$$^\prime$, which is the measured beamsize of Hi-GAL observations at 500 $\mu$m (Traficante et al. 2011). For each pixel, we use equation
\begin{equation}
I_\nu = B_\nu (1 - e^{-\tau_\nu })
\end{equation}
to model intensities at various wavelengths. The optical depth $\tau_\nu$ could be estimated through
\begin{equation}
\tau_\nu = \mu_{H_2} m_H \kappa_\nu N_{H_2} / R_{gd}
\end{equation}
We adopt a mean molecular weight per H$_2$ molecule of $\mu_{H_2}$ = 2.8 to include the contributions from Helium and other heavy elements. $m_H$ is the mass of a hydrogen atom. $N_{H_2}$ is the column density. R$_{gd}$ is the gas-to-dust mass ratio which is set to be 100. According to Ossenkopf $\&$ Henning (1994), dust opacity per unit dust mass ($\kappa_\nu$) could be expressed as
\begin{equation}
\kappa_\nu = 5.0 (\frac{\nu}{600 GHz})^\beta cm^{2} g^{-1}
\end{equation}
where the value of the dust emissivity index $\beta$ is fixed to 1.75 in our fitting. The two free parameters ($N_{H_2}$ and $T_d$) for each pixel could be fitted finally. The final resulting dust temperature and column density maps, which have a spatial resolution of 45$^\prime$$^\prime$ with a pixel size of 13$^\prime$$^\prime$, are shown in Figures 1-6.

MALT90 is an international project with the aim to characterize physical and chemical properties of massive star formation in our Galaxy (e.g., Foster et al. 2011; Foster et al. 2013; Jackson et al. 2013). This project was carried out with the Mopra Spectrometer (MOPS) arrayed on the Mopra 22 m telescope. The beamsize of Mopra is 38$^\prime$$^\prime$ at 86GHz, with a beam efficiency between 0.49 at 86 GHz and 0.42 at 115 GHz (Ladd et al. 2005). The target of this survey are selected from the ATLASGAL clumps found by Contreras et al. (2013). The image size of each MALT90 data cube is about 4$^\prime$$^\prime$ $\times$ 4$^\prime$$^\prime$, with a step of 9$^\prime$$^\prime$. We downloaded the data files from the MALT90 Home Page\footnote{http://atoa.atnf.csiro.au/MALT90}, and assembled all the MALT90 data into a new data cube in a certain region if they have the same velocity component. We have found out six dense molecular clouds showing distinct emissions of HC$_3$N (10-9). Their infrared images and new combined integrated emissions of HC$_3$N (10-9) are also shown in Figures 1-6 respectively. All sources involve at least two ATLASGAL clumps. To calculate the abundance of HC$_3$N in each pixel, we also smoothed the molecular data into a new beamsize of 45$^\prime$$^\prime$ with a new step of 13$^\prime$$^\prime$. By assuming local thermodynamic equilibrium (LTE) conditions and HC$_3$N (10-9) is optically thin, we calculated the HC$_3$N column density in each pixel where its emission is greater than 5 $\sigma$, using the equation from Sanhueza et al. (2012):
\begin{equation}
N = \frac{8 \pi \nu^3}{c^3} \frac{Q_{rot}}{g_u A_{ul}}
\frac{exp(E_l/k T_{ex})}{1 - exp (-h \nu /k T_{ex})} \frac{\int T_{mb} dv}{J(T_{ex}) - J(T_{bg})}
\end{equation}
where $c$ is the velocity of light in the vacuum, $g_u$ is the statistical weight of the upper level, $A_{ul}$ is the Einstein coefficient for spontaneous transition, $E_l$ is the energy of the lower level, $Q_{rot}$ is the partition function, $T_{bg}$ is the background temperature, $T_{ex}$ is the excitation temperature. Like the assumption make by Sanhueza et al. (2012), we here also assume that $T_{ex}$ is equal to the dust temperature derived above. The values of $g_u$, $A_{ul}$ and $E_l$ could be found in the Cologne Database for Molecular Spectroscopy (CDMS) (M\"{u}ller et al. 2001, 2005 ). $J(T)$ is defined by
\begin{equation}
J(T) = \frac{h \nu}{k} \frac{1}{e^{h \nu/k T} - 1}
\end{equation}
For the uncertainties of column density, here we only consider the errors from its integrated intensities. We should mention here that to calculate the column density of HC$_3$N, we made an assumption that the HC$_3$N (10-9) line is optically thin. The true column density derived by Eq. (4) should be multiplied by a factor of $\tau$ / (1 - $e$$^{-\tau}$). For an intermediately optically thick line ($\tau$: 0.5 $\sim$ 2), the true column density will be higher by a factor of 1.3 to 2.3 for our sources. The abundance value of HC$_3$N ($\chi$(HC$_3$N)) for each pixel can be calculated through $\chi$(HC$_3$N) = $N$(HC$_3$N)/$N$(H$_2$) finally. The HC$_3$N abundance maps for each source are shown in Figures 7-12.

\section{Individual sources and Discussions}
\subsection{G5.899-0.429}
The dense cloud G5.899-0.429 involves 5 ATLASGAL clumps. Four of
them have been observed by MALT90 (Figure 1, the green boxes). The
distance of this cloud is about 2.9 kpc (Sato et al. 2014). From
Figure 1, we can see that the HC$_3$N (10-9) emission is very
compact and comes from the densest part of this cloud. The most
massive clump AGAL005.884-00.392, which has the strongest emission of HC$_3$N,
is also known as an expanding UCHII region W28 A2 (Wood
\& Churchwell 1989). The expanding velocity of this UCHII regions is
about 35 km s$^{-1}$ (Acord et al. 1998). Near infrared observations
indicate the exciting source of this UCHII region is a young O-type
star (Feldt et al. 2003). Zapata et al. (2019) carried out high
angular resolution observations and found an explosive outflow from
this UCHII region. The south-east part of this cloud involves two
ATLASGAL clumps which show relatively weak emissions of HC$_3$N
(10-9). The 90 cm radio continuum emissions from MAGPIS are shown in
yellow contours in the top left panel of Figure 1. We can see that
radio emissions here are more diffuse and larger than that in the
UCHII region, indicating this might be a normal classical HII
region. From Figure 7, we can see that compared to the south-east
part, $\chi$(HC$_3$N) is more abundant in the UCHII region, and the
abundance of HC$_3$N increases with dust temperature monotonously in
the whole region. The spectra of HCO$^+$ (1-0) shows the so-called ``blue
profile" with extended wing emissions where the $\chi$(HC$_3$N) is
highest, indicating star-forming activities such as infall and outflow.
This result suggests that HC$_3$N prefers to be formed in warm gas
with massive star-forming activities.

\subsection{G12.804-0.199}
This dense molecular cloud is associated with the well-known GMC
W33. The distance of W33 is about 2.4 kpc (Immer et al. 2013). It
includes three large dust clumps (W33 Main, W33 A and W33 B) and
three smaller clumps (W33 Main1, W33 A1 and W33 B1). Even though
these clumps are involved in a whole star-forming complex, radio
line observations found W33 Main and W33 A have a radial velocity of
$\sim$ 36 km s$^{-1}$, while W33 B has a different radial velocity of
$\sim$ 58 km s$^{-1}$. Using molecular line data from MALT90, we
checked both the two velocity components, only founding HC$_3$N
(10-9) emissions in W33 Main and W33 A (see the top right panel of
Figure 2). The 90 cm radio continuum emissions from MAGPIS in Figure
2 show W33 Main as a compact source, which is also known to be an
UCHII region (Keto $\&$ Ho 1989). On the south-east of W33 Main,
there is a strong arc-shaped radio emission. Ho et al. (1986)
suggest this is an ionization front penetrating W33 Main. The
HCO$^+$ (1-0) spectra in W33 Main also shows the so-called ``blue
profile" with extended wing emissions (Figure 8), indicating infall
and outflow activities in W33 Main. We do not found radio emission
in the center of W33 A. However, van der Tak $\&$ Menten (2005)
found faint 43 GHz radio emissions in W33 A with higher resolution
observations. They suggest the faint emissions come from an ionised
wind or a hypercompact HII region (HCHII) in W33 A. Immer et al.
(2014) detected a large number of simple and complex molecules in
W33 A. They suggest W33 A may be in the transition from the hot core
stage to the HCHII region phase. Thus W33 Main is more evolved than
W33 A. From Figure 8, we can see that HC$_3$N is more abundant in
W33 Main than that in W33 A. Like that found in G5.899-0.429, the
abundance of HC$_3$N also increases with dust temperature in
G12.804-0.199. This result also suggests that HC$_3$N prefers to be formed in warm gas
with massive star-forming activities.

\subsection{G326.641+0.612 (RCW 95)}
The dense molecular cloud G326.641+0.612 is associated with the
classical HII region RCW 95 (Rodgers et al. 1960). The kinematic
distance of RCW 95 is about 2.4 kpc (Giveon et al. 2002). YXW18
found that in this region the abundance of N$_2$H$^+$ reaches a
plateau as the dust temperature is above 27 K (see their Figure 10).
They thus suggest the destruction of N$_2$H$^+$ by CO or UV photons
around this classical HII region. From Figure 9, we can see that as
the dust temperature gets hot, the abundance of HC$_3$N also seems
to reach a plateau, indicating the destruction of HC$_3$N by UV
photons.

\subsection{G327.293-0.579 (RCW 97)}
The dense molecular cloud G327.293-0.579 is associated with a
luminous photon dominated region (PDR) around the classical HII
region RCW 97 on the north side, and an IRDC on the south side
(Wyrowski et al. 2006). From Figure 4, we can see that the emission
of HC$_3$N (10-9) mainly comes from the IRDC which hosts the hot
core G327.3-0.6 (Gibbet al. 2000) and extended green object (EGO)
candidate G327.30-0.58 (Cyganowski et al. 2008). Assuming a
kinematic distance of 3.0 kpc (Russeil 2003), the Lyman continuum
flux from RCW97 will be more than 10$^{50}$ photons s$^{-1}$ (Conti
$\&$ Crowther 2004), indicating this is a giant HII region. Both
infall and outflow activities have been found by Leurini et al.
(2017) in the IRDC. From Figure 10, it can be noticed that the
abundance of HC$_3$N is highest in the IRDC, and begins to drop as
the dust temperature gets hotter than 30 K. This is also consistent
with the scenario that HC$_3$N could be destroyed by UV photons.

\subsection{G337.916-0.477 (S36)}
This dense molecular cloud is associated with the infrared bubble
S36 (Churchwell et al. 2006). The radio continuum emissions from
SUMSS shown in Figure 5 indicate this is also a classical HII
region. YXW18 found that in this region, the abundance of N$_2$H$^+$ increases with
dust temperature when it is below 28 K, and then decreases quickly
in the PDR where T$_d$ is hotter than 28 K (also see their Figure
10). From Figure 11, we can see that the situation of HC$_3$N is
similar to that of N$_2$H$^+$ in this region, indicating UV photons
are also destroying HC$_3$N and N$_2$H$^+$ on the PDR of S36.

\subsection{G345.448+0.314}
The dense molecular cloud G345.448+0.314 involves two ATLASGAL
clumps. The two clumps are associated with IRAS 17008-4040 and IRAS
17009-4042 respectively. The radio continuum emissions from SUMSS in
Figure 6 indicate HII regions in these two clumps. Using high
resolution archival data from the Giant Metrewave Radio Telescope
(GMRT), Dewangan et al. (2018) found 13 HII regions with radius in
the range of 0.06 pc and 0.25 pc in these two IRAS sources. The
radius of these HII regions indicate they are still in the UCHII
stage. Like those found in G5.899-0.429 and G12.804-0.199 introduced
above, the abundance of HC$_3$N also increases with dust temperature
(Figure 12), indicating the production of HC$_3$N is more efficient
than its destruction here. This may be because compared to classical
HII regions, UCHII regions are still surrounded by dense gas,
providing shielding against UV radiation. The spectra of HCO$^+$
(1-0) shows red and blue profiles with wing emissions in IRAS
17008-4040 and IRAS 17009-4042, indicating star-forming activities in this
dense molecular cloud. Like that found in the other UCHII regions (G5.899-0.429 and G12.804-0.199) above,
this result also suggests that HC$_3$N prefers to be formed in warm gas
with massive star-forming activities.

\subsection{Discussions}
We found that in dense molecular clouds associated with classical HII regions (RCW 95, RCW 97 and infrared bubble S36), the abundance of HC$_3$N does not increase with dust temperature monotonously. It begins to decrease or reaches a plateau as the dust temperature gets hot. In a previous paper (Yu $\&$ Xu 2016), we also found that the abundance of HC$_3$N decreases with Lyman continuum flux. These studies indicate that HC$_3$N can be destroyed by UV radiation. Chemical network from KIDA\footnote{http://kida.obs.u-bordeaux1.fr/} (Wakelam et al. 2014) tells us that HC$_3$N could be destroyed through reaction:HC$_3$N + Photon $\longrightarrow$ CN + C$_2$H and/or HC$_3$N + Photon $\longrightarrow$ HC$_3$N$^+$ + e$^-$. Besides, UV photons could also destroy C$_2$H$_2$ which is the main progenitor of HC$_3$N through reaction: C$_2$H$_2$ + Photon $\longrightarrow$ C$_2$H + H, leading to the production of HC$_3$N ineffective.

On the other hand, the situation was quite different in UCHII regions of G5.899-0.429, G12.804-0.199 and G345.448+0.14. We found that in these regions, the abundance of HC$_3$N increases with dust temperature. This may be because that in warm gas the progenitors of HC$_3$N (such as C$_2$H$_2$ and CH$_4$) could be easily released into gas phase. Yu $\&$ Wang (2015) found that in massive young stellar objects (MYSOs), the line widths of HC$_3$N are comparable to those of N$_2$H$^+$, which is regarded as a good tracer of cold dense gas.
However, in UCHII regions the line widths of HC$_3$N become broader than those of N$_2$H$^+$. Taniguchi et al. (2018) also found that the line widths of HC$_3$N are significantly broader than those of HC$_5$N. These studies indicate HC$_3$N prefers to exist in more active star-forming regions. From Figure 7, 8, 10 and 12, we can see that the spectra of HCO$^+$ (1-0) show red and blue profiles with wing emissions where the abundance of HC$_3$N is highest. Previous multi-wavelength observations also indicate shock activities (caused
by infall, outflow and/or expanding HII regions) in these regions. We also checked the spectra of HC$_3$N (10-9) in these regions, and found that in G12.804-0.199 and RCW 97 the spectra of HC$_3$N (10-9) show wing emissions (Figure 13). This suggests HC$_3$N might be an outflow shock origin species. The abundance of HC$_3$N could be increased in the passage of shocks. The chemical model of Mendoza et al. (2018) shows that the abundance of HC$_3$N could be directly increased due to mantle sputtering due to the passage of shocks. Besides, their model also indicates shock activities could increase the reaction efficiency of CN with C$_2$H$_2$: C$_2$H$_2$ + CN $\longrightarrow$ HC$_3$N + H.

The velocity dispersion of HC$_3$N (10-9) caused by thermal motions could be estimated by:
\begin{equation}
\Delta V_{therm} = \sqrt{8ln2kT_d(\frac{1}{m_{HC_3N}}+\frac{1}{m})}
\end{equation}
where $m_{HC_3N}$ is the mass of HC$_3$N (51 per amu), $m$ is the mean molecular mass (2.3 per amu). Figure 13 shows the relationship between the abundance of HC$_3$N and its nonthermal line widths ($\Delta$ V$_{nontherm}^2$ $\equiv$ $\Delta$ V$_{FWHM}^2$ - $\Delta$ V$_{therm}^2$) in G12.804-0.199 and RCW 97. It can be noticed that the abundance of HC$_3$N increases with its nonthermal line width in these two regions. This result also suggests that HC$_3$N could be efficiently formed by massive star formation activities. We thus regard that HC$_3$N prefers to be formed in warm gas with massive star-forming activities. We suggest more line observations with higher resolutions to be carried out to found out the chemical evolution of HC$_3$N in massive star-forming regions.

Previous studies of low-mass star-forming regions suggest the ratio of $N$(nitrogen-bearing species)/$N$(carbon-chain species) increases as a cloud evolves (e.g. Suzuki et al. 1992; Benson et al. 1998; Hirota et al. 2009). Suzuki et al. (1992) carried out observations of CCS, HC$_3$N, HC$_5$N and NH$_3$ toward 49 dark cloud cores in the Taurus and Ophiuchus regions. They found carbon-chain molecules like CCS are abundant in the early stages of chemical evolution, while NH$_3$ is abundant in the later stages. They suppose that in the early stages of star formation, carbon-chain molecules could be efficiently formed from ionic carbon (C$^+$) and atomic carbon (C). However, as the cloud evolves further, the formation efficiency of carbon-chain molecules decrease, because most of the carbon atoms are converted into the form of CO, which is chemically stable. On the other hand, nitrogen-bearing species such as NH$_3$ become gradually abundant in the central part of the core. Thus the abundance ratio of CCS and NH$_3$ could be used as a good indicator of cloud evolution and star formation. Benson et al. (1998) made a high spatial resolution observation of N$_2$H$^+$, C$_3$H$_2$ and CCS toward 60 dense cores. They found that the ratio of $N$(CCS)/$N$(N$_2$H$^+$) in starless cores is higher by a factor of 2 than that in cores with stars. This result is consistent with the finding of Suzuki et al. (1992).
Ohashi et al. (2014; 2016) have investigated molecular lines of HC$_3$N (10-9) and N$_2$H$^+$ (1-0) in the cluster forming regions of Orion A and Vela C GCMs. They found that the ratios of $N$(NH$_3$)/$N$(CCS) and $N$(N$_2$H$^+$)/$N$(HC$_3$N) could also be the tracers of the chemical evolution even in the high mass star forming region in the same way as the low mass star forming region. Figure 14 shows the relationship between $N$(N$_2$H$^+$)/$N$(HC$_3$N) and $T_{dust}$ in all of our sources. It can be noticed that in all sources, the relative column density ratios of N$_2$H$^+$ and HC$_3$N do not increase with dust temperature. Moreover, in G5.899-0.429, G12.804-0.199, RCW 97, S36 and G345.448+0.314, it is clear that this ratio decreases as dust temperature increases, which is totally different to that found by Ohashi et al. (2016).
The reason may be that we focused on the different evolutionary stages of massive star formation. It is generally accepted that massive stars evolve from starless cores in IRDCs to hot cores with central young stellar objects, then to HCHII and UCHII regions. The final stages are compact and classical HII regions (Zinnecker et al. 2007). Previous studies of Ohashi et al. (2016) compared starless cores with star-forming (Class 0/I) sources. In this work, we focused on sources of UCHII regions and classical HII regions, where massive protostars have already formed. According to the new chemistry of WCCC, Sakai et al. (2008) suggested that the HC$_3$N will be more enhanced after star formation starts due to the evaporation from the grain surface. The increase of $\chi$(HC$_3$N) with dust temperature has been founded in G5.899-0.429 (Figure 7), G12.804-0.199 (Figure 8) and G345.448+0.314 (Figure 12), and the decrease of $\chi$(N$_2$H$^+$) with dust temperature in S36 has also been shown in our previous paper of YXW18. This may be the reason that the ratio of $N$(N$_2$H$^+$)/$N$(HC$_3$N) decreases with the dust temperature in our sources. Our study suggests this ratio still could be used as a chemical evolutionary indicator of cloud evolution after the massive star formation is started.

\section{Summary}
We investigate the chemical evolution of HC$_3$N in six dense molecular clouds, using data from MALT90 and Hi-GAL. Radio sky surveys indicate these dense molecular clouds are associated with UCHII regions and/or classical HII regions. We found that in dense molecular clouds associated with classical HII regions, the abundance of HC$_3$N decreases or reaches a plateau when the dust temperature gets hot, implying UV photons could destroy the molecule of HC$_3$N. On the other hand, in dense molecular clouds associated with UCHII regions, we found the abundance of HC$_3$N increases with dust temperature monotonously. The spectra of HCO$^+$ (1-0) and HC$_3$N (10-9) in some sources show wing emissions. We also found that the abundance of HC$_3$N increases with its nonthermal velocity width in G12.804-0.199 and RCW 97. These results suggest HC$_3$N prefers to be formed in warm gas with star-forming activities, and could be destroyed by UV photons in the late stages of massive star formation. We also found that in most sources, the column density ratio of N$_2$H$^+$ and HC$_3$N decreases with the dust temperature. Our study seems to support that the column density ratio of N$_2$H$^+$ and HC$_3$N could still be used as a chemical evolutionary indicator of cloud evolution after the massive star formation is started.

\section{ACKNOWLEDGEMENTS}
We thank the anonymous referee for constructive suggestions.
This paper has made use of information from the ATLASGAL Database
Server\footnote{
http://atlasgal.mpifr-bonn.mpg.de/cgi-bin/ATLASGAL$_-$DATABASE.cgi}.
The ATLASGAL project is a collaboration between the
Max-Planck-Gesellschaft, the European Southern Observatory (ESO) and
the Universidad de Chile. This research made use of data products
from the Millimetre Astronomy Legacy Team 90 GHz (MALT90) survey.
The Mopra telescope is part of the Australia Telescope and is funded
by the Commonwealth of Australia for operation as National Facility
managed by CSIRO. This paper is supported by the Youth Innovation Promotion
Association of CAS.


\begin{thebibliography}{h}
\bibitem[\protect\citeauthoryear{Contreras et al.}{2007a}]{b21} Acord J. M., Churchwell E., Wood D. O. S. 1998, ApJ, 495, L107

\bibitem[\protect\citeauthoryear{Contreras et al.}{2007a}]{b21} Benson P. J., Caselli P., \& Myers P. C. 1998, ApJ, 506, 743

\bibitem[\protect\citeauthoryear{Contreras et al.}{2007a}]{b21} Cernicharo J., Gu$\acute{e}$lin M. 1996, A\&A, 309, L27

\bibitem[\protect\citeauthoryear{Contreras et al.}{2007a}]{b21} Chapman J. F., Millar T. J., Wardle M., Burton M. G., Walsh A. J. 2009, MNRAS, 394, 221

\bibitem[\protect\citeauthoryear{Contreras et al.}{2007a}]{b21} Churchwell, E., Povich, M. S., Allen, D., et al. 2006, ApJ, 649, 759

\bibitem[\protect\citeauthoryear{Contreras et al.}{2007a}]{b21} Conti P. S., Crowther P. A. 2004, MNRAS, 355, 899

\bibitem[\protect\citeauthoryear{Contreras et al.}{2007a}]{b21} Contreras Y., Schuller F., Urquhart J. S., et al. 2013, A\&A, 549, A45

\bibitem[\protect\citeauthoryear{Contreras et al.}{2007a}]{b21} Crovisier J. et al. 2004, A\&A, 418, 1141

\bibitem[\protect\citeauthoryear{Contreras et al.}{2007a}]{b21} Cyrowski C. J., Whitney B. A., Holden E. et al. 2008, AJ, 136, 2391

\bibitem[\protect\citeauthoryear{Contreras et al.}{2007a}]{b21} Dewangan L. K., Baug T., Ojha D. K., Ghosh S. K. 2018, ApJ, 869, 30

\bibitem[\protect\citeauthoryear{Contreras et al.}{2007a}]{b21} Feldt M., Puga, E., Lenzen, R. et al. 2003, ApJ, 599, L91

\bibitem[\protect\citeauthoryear{Contreras et al.}{2007a}]{b21} Foster J. B., Jackson J. M., Barnes P. J. et al. 2011, ApJS, 197, 25

\bibitem[\protect\citeauthoryear{Contreras et al.}{2007a}]{b21} Gibb E., Nummelin A., Irvine W. M., Whittet D. C. B., Bergman P. 2000, ApJ, 545, 309

\bibitem[\protect\citeauthoryear{Contreras et al.}{2007a}]{b21} Giveon A., Sternberg A., Lutz D., Fruchtgruber H., Pauldrach, A. W. A. 2002, ApJ, 566, 880

\bibitem[\protect\citeauthoryear{Contreras et al.}{2007a}]{b21} Hassel G. E., Herbst E., Garrod R. T. 2008, ApJ, 681, 1385

\bibitem[\protect\citeauthoryear{Contreras et al.}{2007a}]{b21} Hirota T. Ohishi M. \& Yamamoto S. 2009, ApJ, 699, 585

\bibitem[\protect\citeauthoryear{Contreras et al.}{2007a}]{b21} Ho P. T. P., Klein R. I., Haschick A. D. 1986, ApJ, 305, 714

\bibitem[\protect\citeauthoryear{Contreras et al.}{2007a}]{b21} Immer K., Reid M. J., Menten K. M., Brunthaler A., Dame T. M. 2013, A\&A, 553, A117

\bibitem[\protect\citeauthoryear{Contreras et al.}{2007a}]{b21} Immer K., Galv$\acute{a}$n-Madrid R., K$\ddot{o}$nig C., Liu H. B., Menten K. M. 2014, A\&A, 572, A63

\bibitem[\protect\citeauthoryear{Contreras et al.}{2007a}]{b21} Jackson, J. M., Rathborne, J. M., Foster, J. B., et al. 2013, PASA, 30, 57

\bibitem[\protect\citeauthoryear{Contreras et al.}{2007a}]{b21}  Keto E. R., \& Ho P. T. P. 1989, ApJ, 347, 349

\bibitem[\protect\citeauthoryear{Contreras et al.}{2007a}]{b21} Ladd, N., Purcell, C., Wong, T., \& Robertson, S. 2005, PASA, 22, 62

\bibitem[\protect\citeauthoryear{Contreras et al.}{2007a}]{b21} Lada C. J., Lombardi M., \& Alves J. F. 2010, ApJ, 724, 687

\bibitem[\protect\citeauthoryear{Contreras et al.}{2007a}]{b21} Leurini S., Herpin F., van der Tak F., Wrowski F., Herczeg G. J., van Dishoeck E. F. 2017, A\&A, 602, A70

\bibitem[\protect\citeauthoryear{Contreras et al.}{2007a}]{b21} Mendoza E., Lefloch B., Ceccarelli C. et al. 2018, MNRAS, 475, 5501

\bibitem[\protect\citeauthoryear{Contreras et al.}{2007a}]{b21} Millar T. J., Macdonald G. H., Gibb A.  G. 1997, A\&A, 325, 1163

\bibitem[\protect\citeauthoryear{Lumsden et al.}{2007a}]{b21} M\"{u}ller H. S. P., Thorwirth S., Roth D. A., \& Winnewisser G.
2001, A\&A, 370, L49
\bibitem[\protect\citeauthoryear{Lumsden et al.}{2007a}]{b21} M\"{u}ller H. S. P., Schl\"{o}der F., Stutzki J., \& Winnewisser G. 2005, J. Molec. Struct., 742, 215

\bibitem[\protect\citeauthoryear{Lumsden et al.}{2007a}]{b21} Ohashi S., Tatematsu K., Choi M., et al. 2014, PASJ, 66, 119

\bibitem[\protect\citeauthoryear{Lumsden et al.}{2007a}]{b21} Ohashi S., Tatematsu K., Fujii K., et al. 2016, PASJ, 68, 3

\bibitem[\protect\citeauthoryear{Contreras et al.}{2007a}]{b21} Ossenkopf V., Henning T., 1994, A\&A, 291, 943

\bibitem[\protect\citeauthoryear{Contreras et al.}{2007a}]{b21} Rathborne J. M., Whitaker J. S., Jackson J. M., et al. 2016, Publ. Astron. Soc. Aust., 33, e030

\bibitem[\protect\citeauthoryear{Contreras et al.}{2007a}]{b21} Rodgers A. W., Campbell C. T., \& Whiteoak J. B. 1960, MNRAS, 121, 103

\bibitem[\protect\citeauthoryear{Contreras et al.}{2007a}]{b21} Russeil D. 2003, A\&A, 397, 133

\bibitem[\protect\citeauthoryear{Contreras et al.}{2007a}]{b21} Sakai N. Sakai T. Hirota T. Yamamoto S. 2008 ApJ, 672, 371

\bibitem[\protect\citeauthoryear{Contreras et al.}{2007a}]{b21} Sakai N., \& Yamamoto S. 2013, ChRv, 113, 8981

\bibitem[\protect\citeauthoryear{Contreras et al.}{2007a}]{b21} Sanhueza, P., Jackson, J. M., Foster, J. B., Garay, G., Silva, A., Finn, S. C., 2012, ApJ, 756, 60

\bibitem[\protect\citeauthoryear{Contreras et al.}{2007a}]{b21} Sato M., Wu Y. W., Immer K., et al. 2014, ApJ, 793, 72

\bibitem[\protect\citeauthoryear{Contreras et al.}{2007a}]{b21} Suzuki H., Yamamoto S., Ohishi M. et al. 1992, ApJ, 392, 551

\bibitem[\protect\citeauthoryear{Contreras et al.}{2007a}]{b21} Taniguchi K., Saito M., Ozeki H. 2016, ApJ, 830, 106

\bibitem[\protect\citeauthoryear{Contreras et al.}{2007a}]{b21} Taniguchi K., Saito M., Sridharan T. K., Minamidani T. 2018, ApJ, 854, 133

\bibitem[\protect\citeauthoryear{Contreras et al.}{2007a}]{b21} Taniguchi K., Saito M., Sridharan T. K., Minamidani T. 2019, ApJ, 872, 154

\bibitem[\protect\citeauthoryear{Contreras et al.}{2007a}]{b21} Takano S., et al. 1998, A\&A, 329, 1156

\bibitem[\protect\citeauthoryear{Contreras et al.}{2007a}]{b21} Traficante, A., Calzoletti, L., Veneziani, M., et al. 2011, MNRAS, 416, 2932

\bibitem[\protect\citeauthoryear{Contreras et al.}{2007a}]{b21} Turner B. E. 1971, ApJ, 163, L35

\bibitem[\protect\citeauthoryear{Contreras et al.}{2007a}]{b21} Urquhart J .S., Figura C., Wyrowski F., et al. 2019, MNRAS, 484, 4444

\bibitem[\protect\citeauthoryear{Contreras et al.}{2007a}]{b21} van der Tak F. F. S., \& Menten K. M. 2005, A\&A, 437, 947

\bibitem[\protect\citeauthoryear{Contreras et al.}{2007a}]{b21} Wakelam V., Vastel C., Aikawa Y., et al. 2014, MNRAS, 445, 2854

\bibitem[\protect\citeauthoryear{Contreras et al.}{2007a}]{b21} Wang, K., Testi, L., Ginsburg, A., et al. 2015, MNRAS, 450, 4043

\bibitem[\protect\citeauthoryear{Contreras et al.}{2007a}]{b21} Wood D. O., Churchwell E. 1989, ApJS, 69, 831

\bibitem[\protect\citeauthoryear{Contreras et al.}{2007a}]{b21} Wyrowski F., Menten K. M., Schilke P., et al. 2006, A\&A, 454, L91

\bibitem[\protect\citeauthoryear{Contreras et al.}{2007a}]{b21} Yamamoto T., Nakagawa N., \& Fukui Y. 1983, A\&A, 122, 171

\bibitem[\protect\citeauthoryear{Contreras et al.}{2007a}]{b21} Yu N. P., \& Wang J. J. 2015, MNRAS, 451, 2507

\bibitem[\protect\citeauthoryear{Contreras et al.}{2007a}]{b21} Yu N. P., \& Xu J. L. 2016, ApJ, 833, 248

\bibitem[\protect\citeauthoryear{Contreras et al.}{2007a}]{b21} Yu N. P., Xu J. L., Wang J. J., Liu X. L. 2018, ApJ, 865, 135

\bibitem[\protect\citeauthoryear{Contreras et al.}{2007a}]{b21} Zapata L. A., Ho P. T. P., Guzm$\acute{a}$n E. et al. 2019, arXiv:1904.04385

\bibitem[\protect\citeauthoryear{Contreras et al.}{2007a}]{b21} Zinnecker H., Yorke H. W., 2007, ARA\&A, 45, 481

\end{thebibliography}

\clearpage

\begin{figure*}
\centering
\includegraphics[width=0.9\textwidth, angle=0]{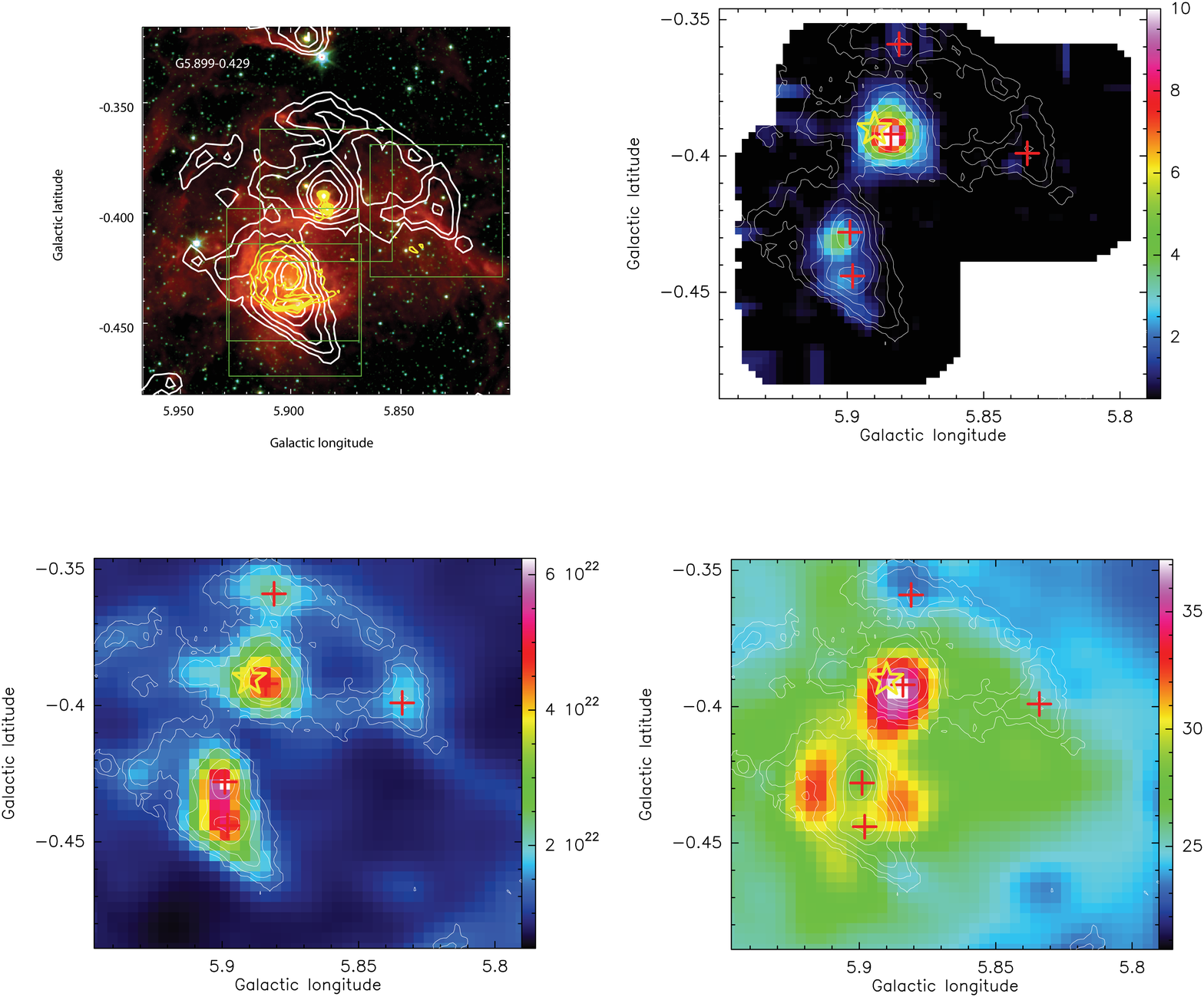}
\caption{Top left:
Three colour mid-infrared image of G5.899-0.429 created using the
Spitzer IRAC band filters (8.0 $\mu$m in red, 4.5 $\mu$m in green
and 3.6 $\mu$m in blue). The green boxes indicate the four observed
regions by MALT90. Top right: The new combined image of HC$_3$N from
MALT90 data set. The emission has been integrated from 3 to 15 km
s$^{-1}$. The unit of the color bar on the right is in K km/s.
Bottom left: The H$_2$ column density of this region built through
SED fitting. The unit of color bar is in cm$^{-2}$. Bottom right:
The dust temperature map in color scale derived from the SED
fitting. The unit of color bar is in K. The ATLASGAL 870 $\mu$m
emissions (in white) are superimposed with levels 0.24, 0.48, 0.96,
1.92 and 3.84 Jy/beam in each panel. The red pluses mark the center
locations of ATLASGAL clumps. The yellow star marks the UCHII region
W28 A2. The 90 cm radio continuum emissions (in yellow) from MAGPIS
are superimposed with levels 0.1, 0.2, 0.3, 0.4, and 0.5 Jy
beam$^{-1}$.}
\end{figure*}

\begin{figure*}
\centering
\includegraphics[width=0.9\textwidth, angle=0]{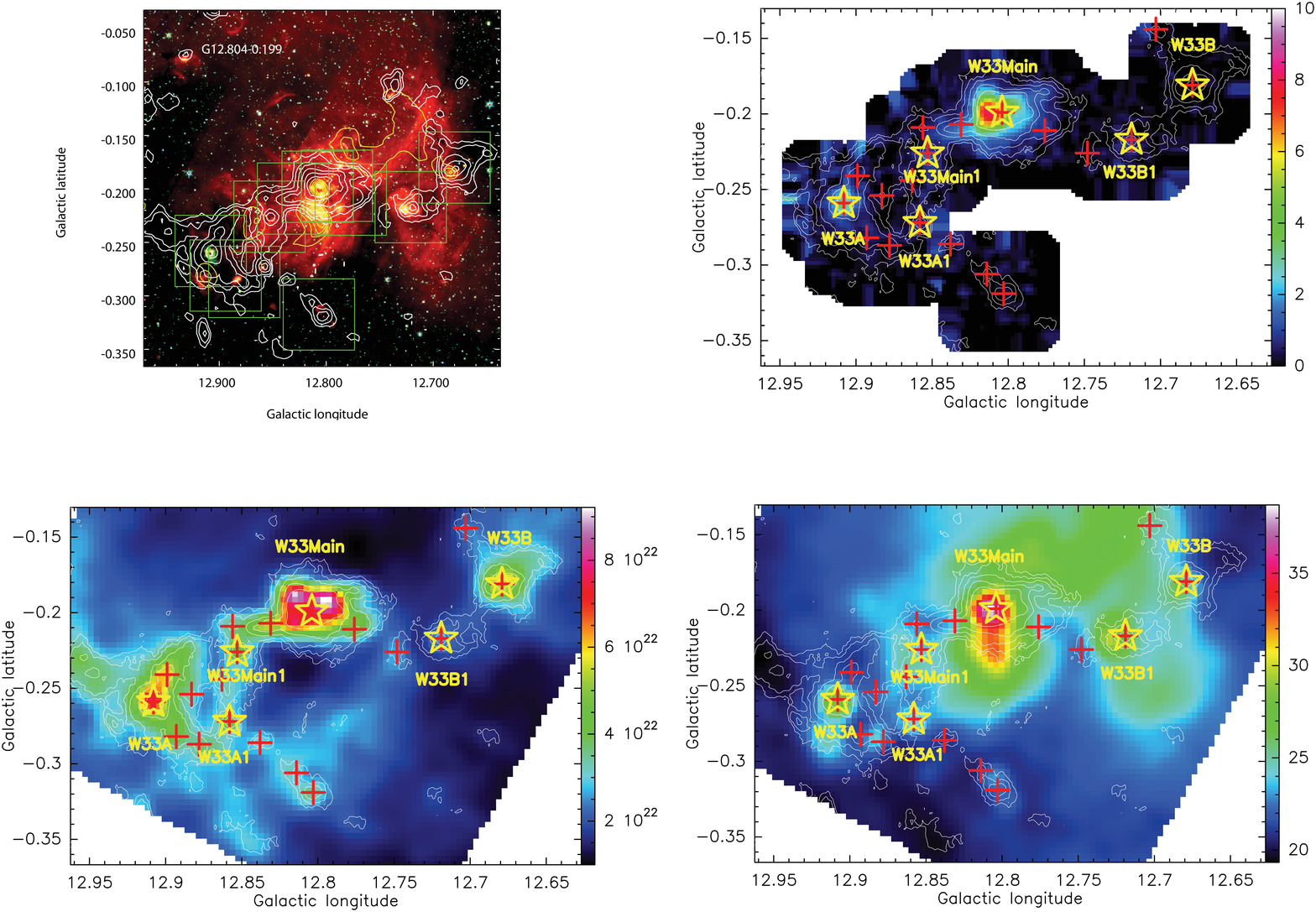}
\caption{Top left:
Three colour mid-infrared image of G12.804-0.199 created using the
Spitzer IRAC band filters (8.0 $\mu$m in red, 4.5 $\mu$m in green
and 3.6 $\mu$m in blue). The green boxes indicate the observed
regions by MALT90. Top right: The new combined image of HC$_3$N from
MALT90 data set. The emission has been integrated from 30 to 40 km
s$^{-1}$. The unit of the color bar on the right is in K km/s.
Bottom left: The H$_2$ column density of this region built through
SED fitting. The unit of color bar is in cm$^{-2}$. Bottom right:
The dust temperature map in color scale derived from the SED
fitting. The unit of color bar is in K. The ATLASGAL 870 $\mu$m
emissions (in white) are superimposed with levels 0.24, 0.48, 0.96,
1.92, 3.84 and 7.68 Jy/beam in each panel. The red pluses mark the
center locations of ATLASGAL clumps. The yellow stars mark W33A,
W33A1, W33Main1, W33Main, W33B1 and W33B. The 90 cm radio continuum
emissions (in yellow) from MAGPIS are superimposed with levels 0.02,
0.04, 0.08, 0.16, and 0.32 Jy beam$^{-1}$. }
\end{figure*}

\begin{figure*}
\centering
\includegraphics[width=0.9\textwidth, angle=0]{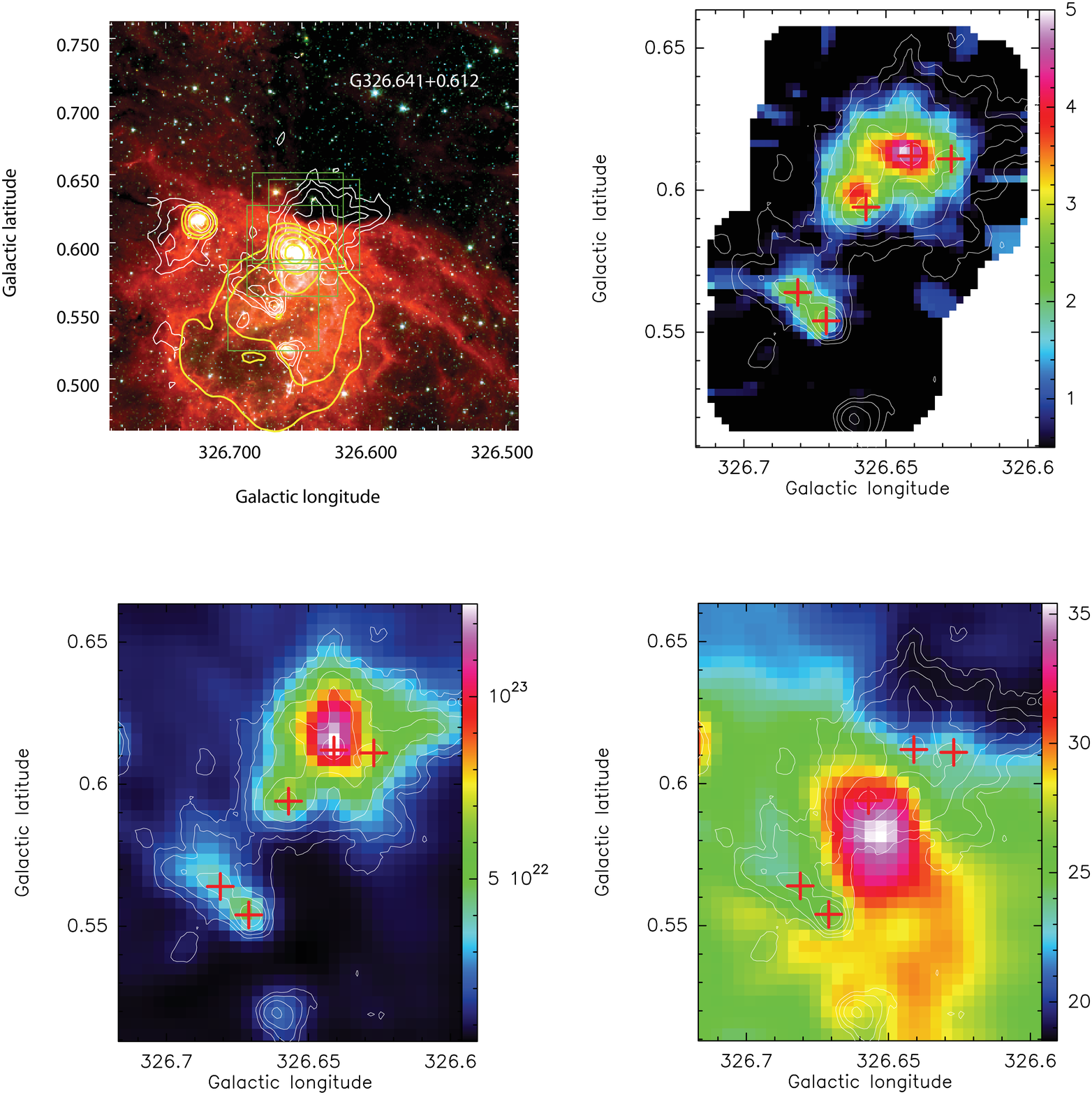}
\caption{Top left:
Three colour mid-infrared image of G326.641+0.612 created using the
Spitzer IRAC band filters (8.0 $\mu$m in red, 4.5 $\mu$m in green
and 3.6 $\mu$m in blue). The green boxes indicate the four observed
regions by MALT90. Top right: The new combined image of HC$_3$N from
MALT90 data set. The emission has been integrated from -44 to -36 km
s$^{-1}$. The unit of the color bar on the right is in K km/s.
Bottom left: The H$_2$ column density of this region built through
SED fitting. The unit of color bar is in cm$^{-2}$. Bottom right:
The dust temperature map in color scale derived from the SED
fitting. The unit of color bar is in K. The ATLASGAL 870 $\mu$m
emissions (in white) are superimposed with levels 0.42, 0.84, 1.68,
3.36, and 6.72 Jy/beam in each panel. The red pluses mark the center
locations of ATLASGAL clumps. The 843 MHz SUMSS radio continuum
emissions (in yellow) are superimposed with levels 0.2, 0.4, 0.8,
1.6, and 3.2 Jy beam$^{-1}$. }
\end{figure*}

\begin{figure*}
\centering
\includegraphics[width=0.9\textwidth, angle=0]{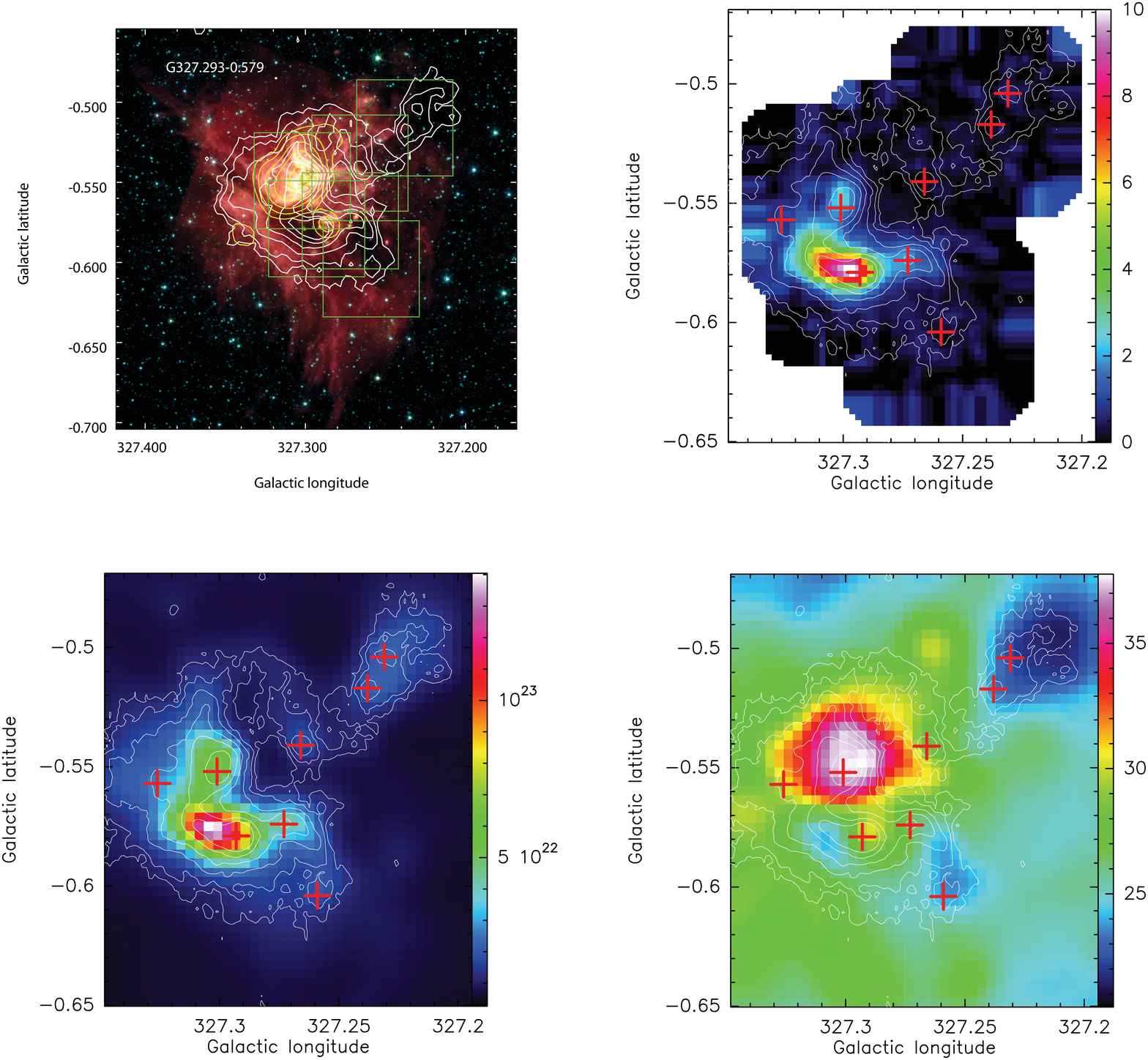}
\caption{Top left:
Three colour mid-infrared image of G327.293-0.579 created using the
Spitzer IRAC band filters (8.0 $\mu$m in red, 4.5 $\mu$m in green
and 3.6 $\mu$m in blue). The green boxes indicate the five observed
regions by MALT90. Top right: The new combined image of HC$_3$N from
MALT90 data set. The emission has been integrated from -50 to -41 km
s$^{-1}$. The unit of the color bar on the right is in K km/s.
Bottom left: The H$_2$ column density of this regions built through
SED fitting. The unit of color bar is in cm$^{-2}$. Bottom right:
The dust temperature map in color scale derived from the SED
fitting. The unit of color bar is in K. The ATLASGAL 870 $\mu$m
emissions (in white) are superimposed with levels 0.21, 0.42, 0.84,
1.68, 3.36, and 6.72 Jy/beam in each panel. The red pluses mark the
center locations of ATLASGAL clumps. The 843 MHz SUMSS radio
continuum emissions (in yellow) are superimposed with levels 0.3,
0.6, 1.2, 2.4, and 4.8 Jy beam$^{-1}$.  }
\end{figure*}

\begin{figure*}
\centering
\includegraphics[width=0.9\textwidth, angle=0]{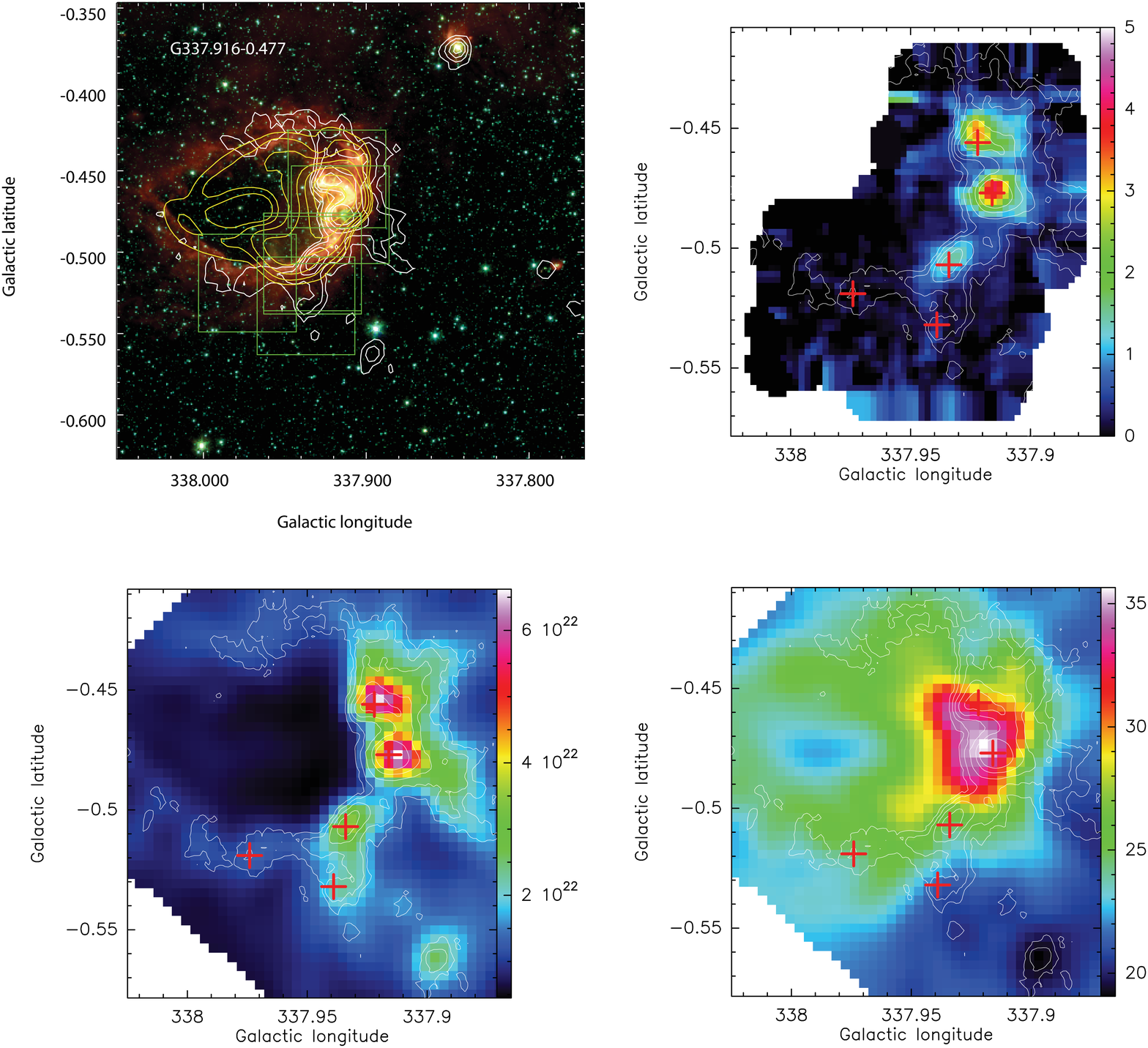}
\caption{Top left:
Three colour mid-infrared image of G337.916-0.477 created using the
Spitzer IRAC band filters (8.0 $\mu$m in red, 4.5 $\mu$m in green
and 3.6 $\mu$m in blue). The green boxes indicate the six observed
regions by MALT90. Top right: The new combined image of HC$_3$N from
MALT90 data set. The emission has been integrated from -42 to -37 km
s$^{-1}$. The unit of the color bar on the right is in K km/s.
Bottom left: The H$_2$ column density of this regions built through
SED fitting. The unit of color bar is in cm$^{-2}$. Bottom right:
The dust temperature map in color scale derived from the SED
fitting. The unit of color bar is in K. The ATLASGAL 870 $\mu$m
emissions (in white) are superimposed with levels 0.27, 0.54, 1.08,
2.16, and 4.32 Jy/beam in each panel. The red pluses mark the center
locations of ATLASGAL clumps. The 843 MHz SUMSS radio continuum
emissions (in yellow) are superimposed with levels 0.05, 0.10, 0.20,
0.40, 0.80, 1.60 and 3.20 Jy beam$^{-1}$. }
\end{figure*}

\begin{figure*}
\centering
\includegraphics[width=0.9\textwidth, angle=0]{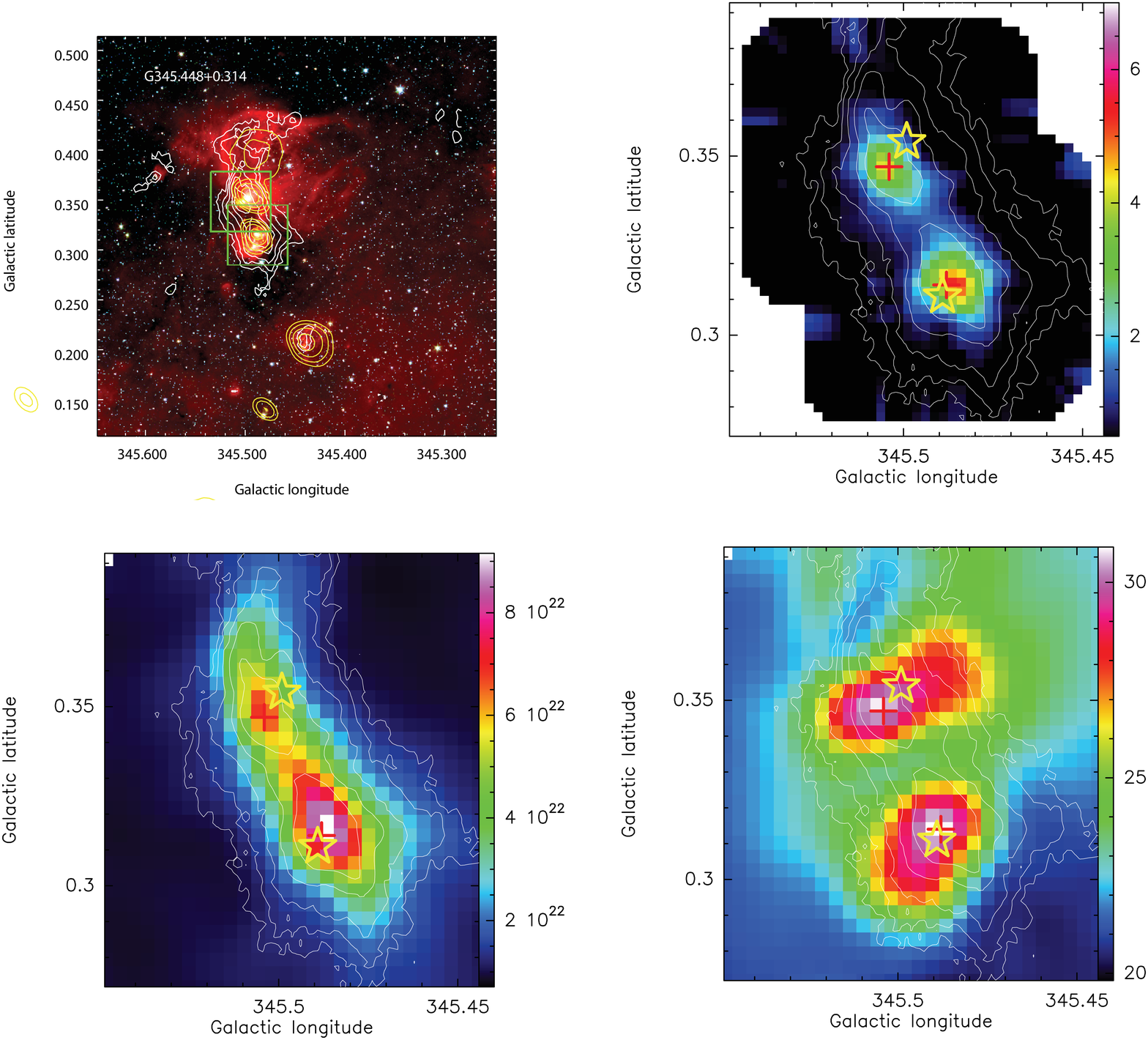}
\caption{Top left:
Three colour mid-infrared image of G345.448+0.314 created using the
Spitzer IRAC band filters (8.0 $\mu$m in red, 4.5 $\mu$m in green
and 3.6 $\mu$m in blue). The green boxes indicate the two observed
regions by MALT90. Top right: The new combined image of HC$_3$N from
MALT90 data set. The emission has been integrated from -21 to -14 km
s$^{-1}$. The unit of the color bar on the right is in K km/s.
Bottom left: The H$_2$ column density of this regions built through
SED fitting. The unit of color bar is in cm$^{-2}$. Bottom right:
The dust temperature map in color scale derived from the SED
fitting. The unit of color bar is in K. The ATLASGAL 870 $\mu$m
emissions (in white) are superimposed with levels 0.18, 0.36, 0.72,
1.44, and 2.88 Jy/beam in each panel. The red pluses mark the center
locations of ATLASGAL clumps. The two yellow stars mark IRAS
17008-4040 and IRAS 17009-4042. The 843 MHz SUMSS radio continuum
emissions (in yellow) are superimposed with levels 0.02, 0.04, 0.08,
and 0.16 Jy beam$^{-1}$. }
\end{figure*}

\begin{figure*}
\centering
\psfig{file=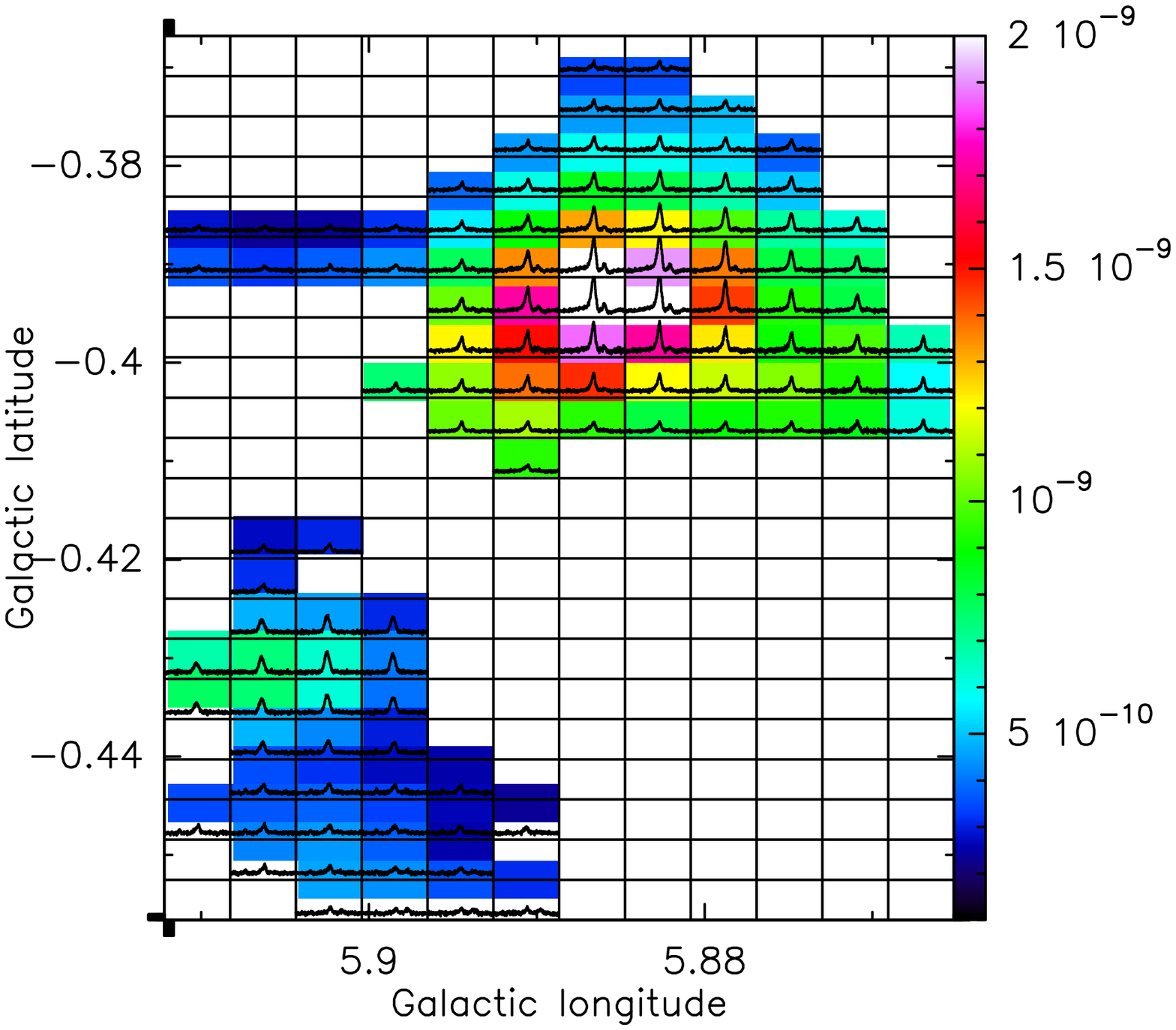,width=3.5in,height=3.3in}
\psfig{file=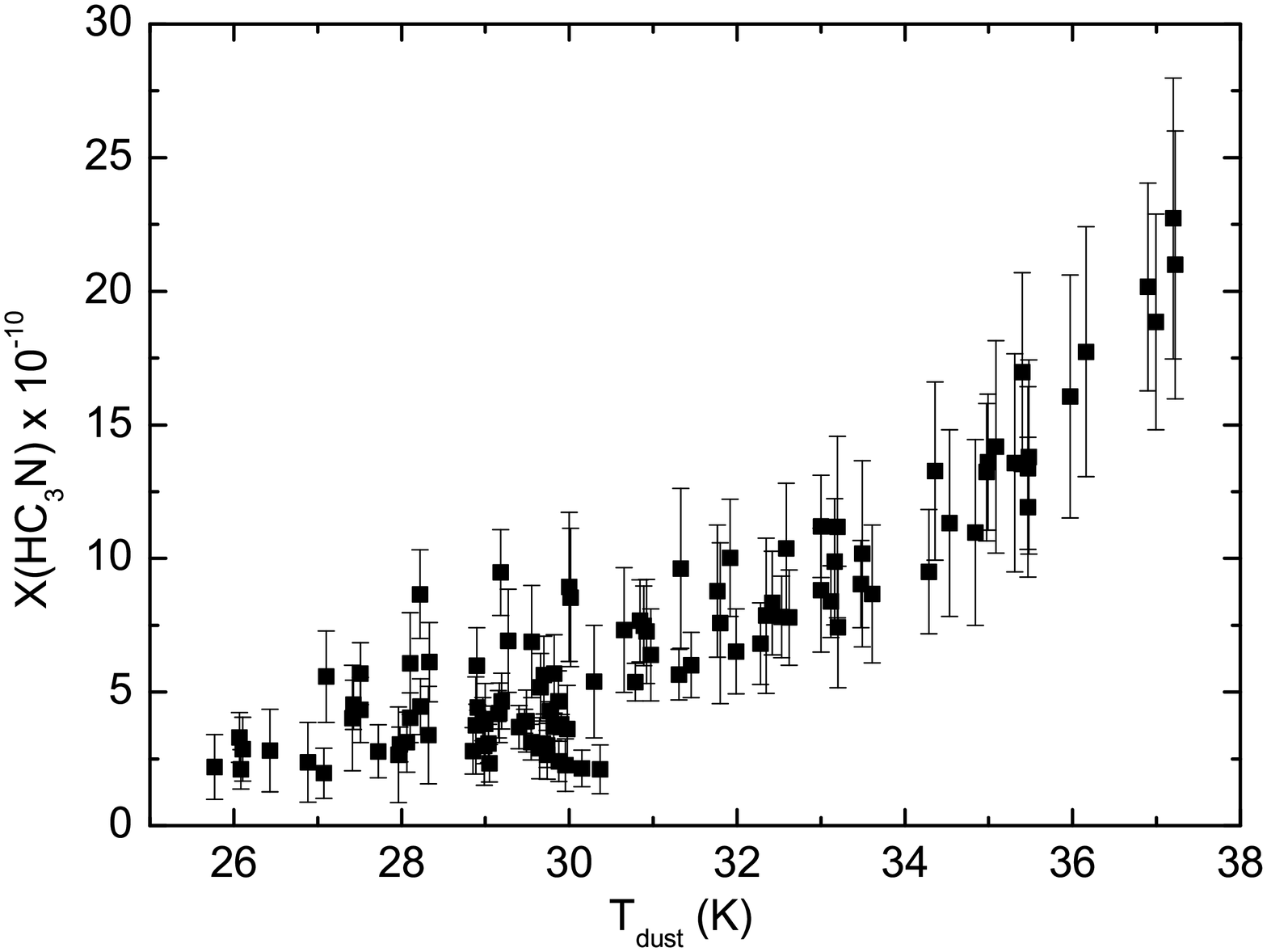,width=4.5in,height=3.3in} \caption{Top:
The HCO$^+$ (1-0) spectra superimposed on the calculated HC$_3$N
abundance map of G5.899-0.429. Bottom: Abundance of HC$_3$N plotted
as a function of the dust temperature in each pixel of G5.899-0.429.
}
\end{figure*}

\begin{figure*}
\centering
\psfig{file=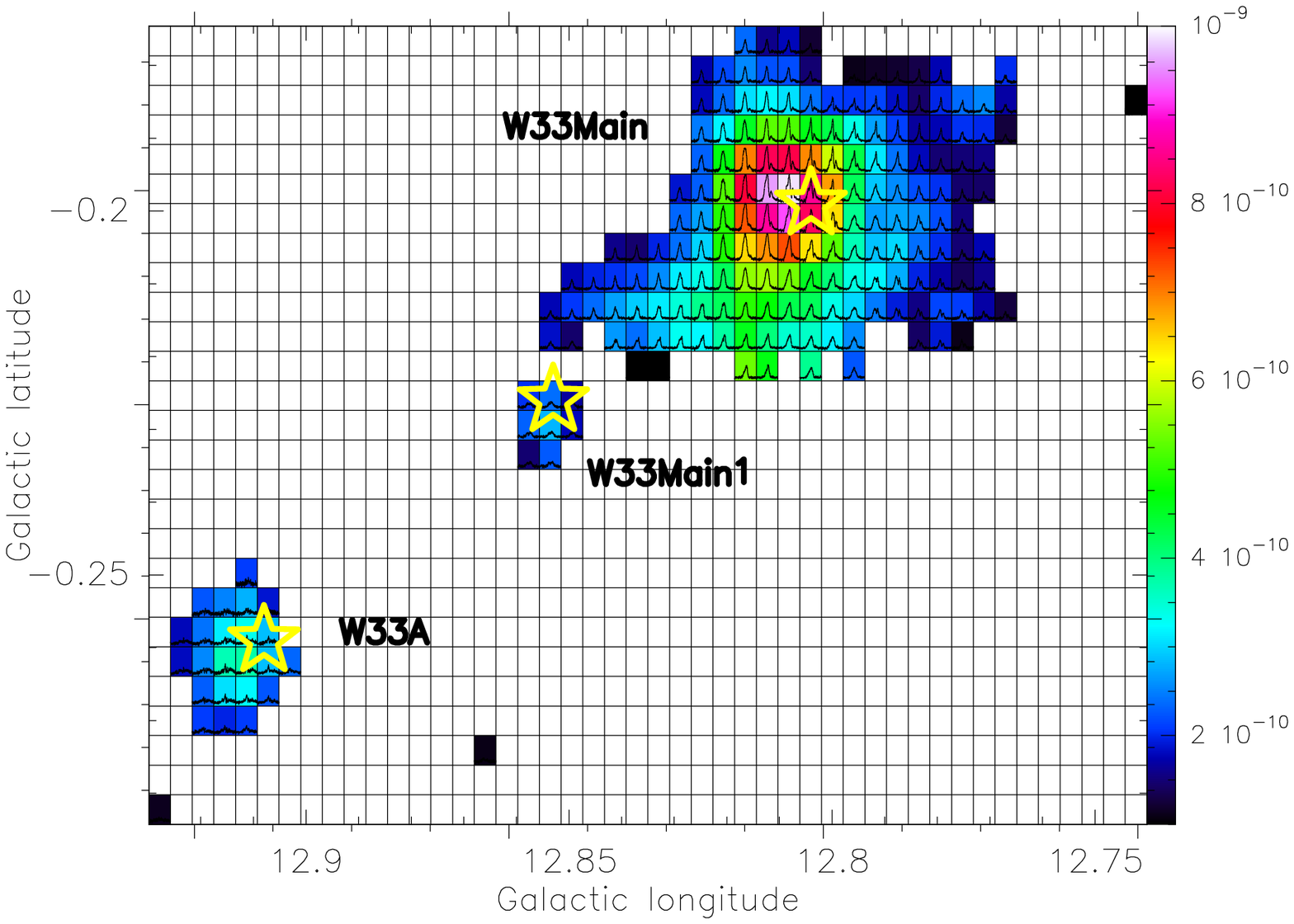,width=4.3in,height=3.3in}
\psfig{file=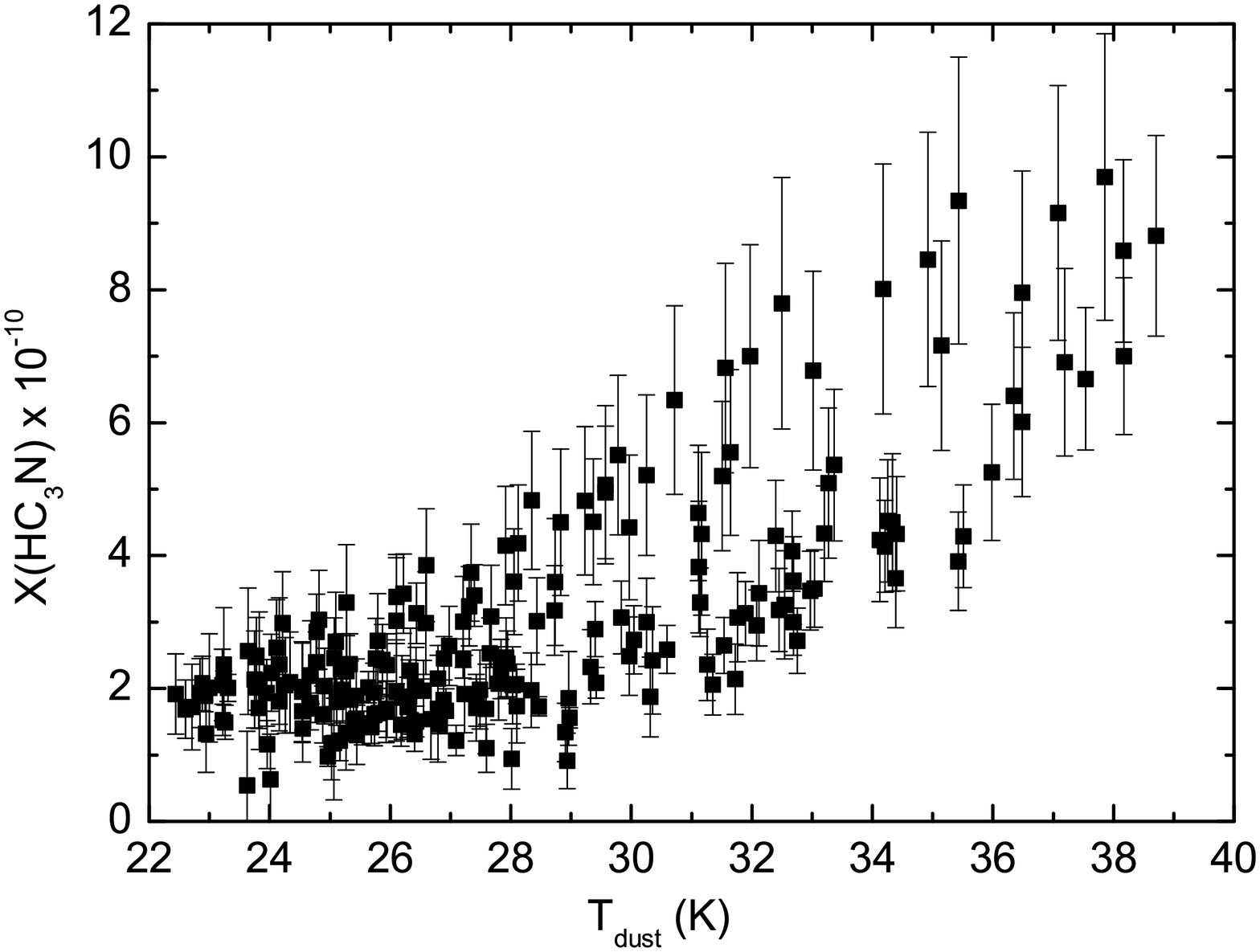,width=4.5in,height=3.3in} \caption{Top:
The HCO$^+$ (1-0) spectra superimposed on the calculated HC$_3$N
abundance map of G12.804-0.199. Bottom: Abundance of HC$_3$N plotted
as a function of the dust temperature in each pixel of
G12.804-0.199. }
\end{figure*}

\begin{figure*}
\centering
\psfig{file=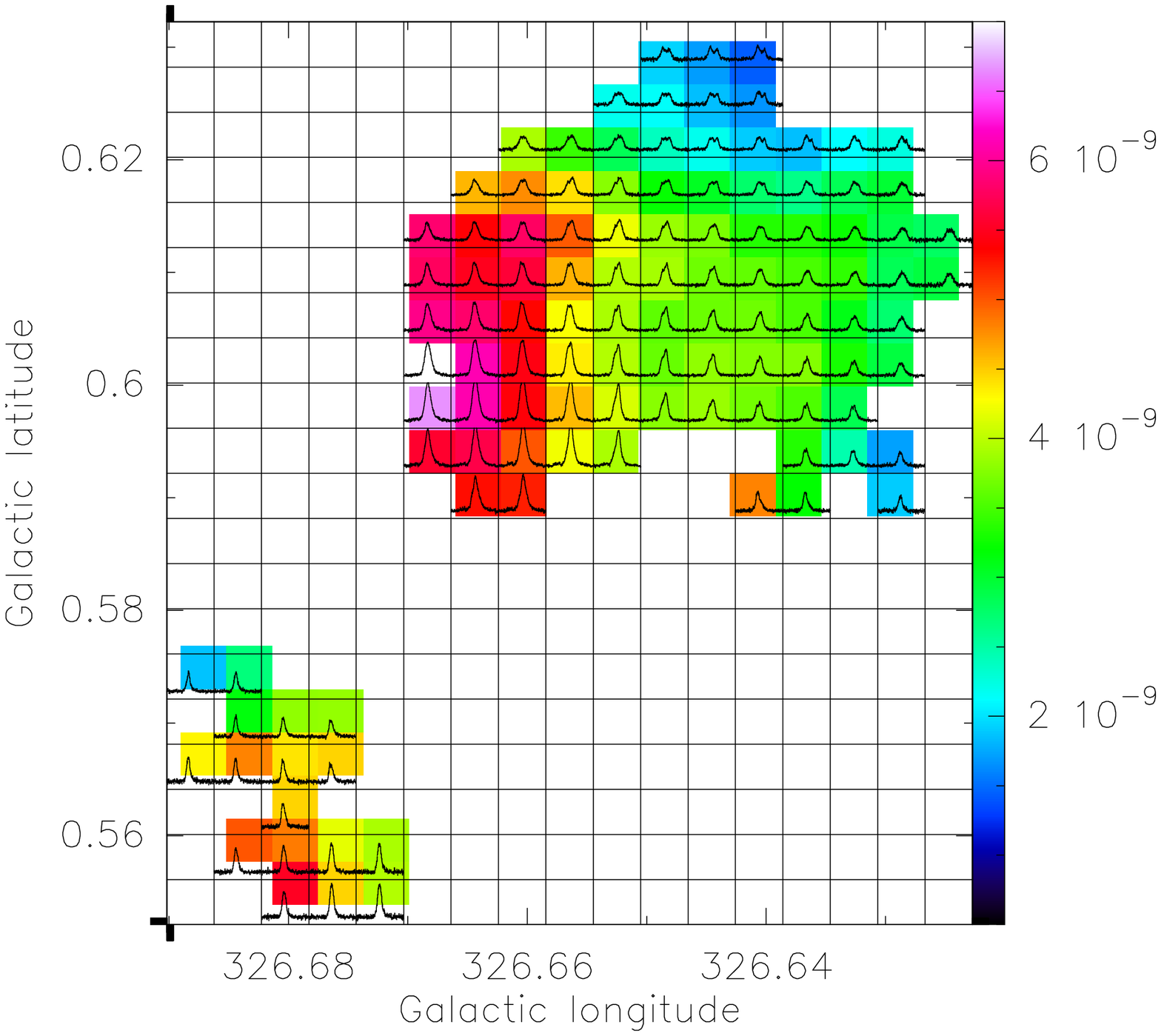,width=3.5in,height=3.3in}
\psfig{file=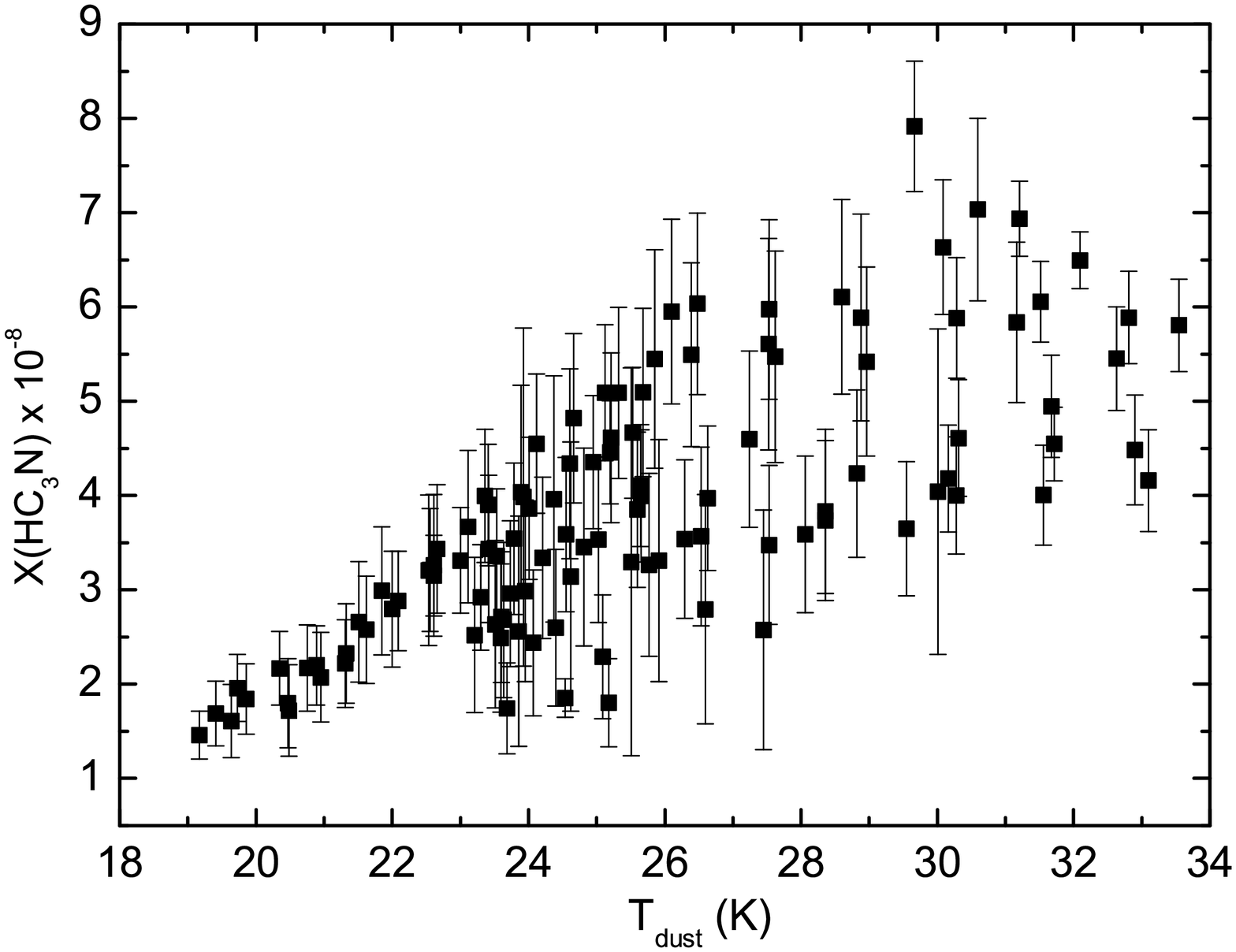,width=4.5in,height=3.3in} \caption{Top:
The HCO$^+$ (1-0) spectra superimposed on the calculated HC$_3$N
abundance map of G326.641+0.612. Bottom: Abundance of HC$_3$N
plotted as a function of the dust temperature in each pixel of
G326.641+0.612. }
\end{figure*}

\begin{figure*}
\centering
\psfig{file=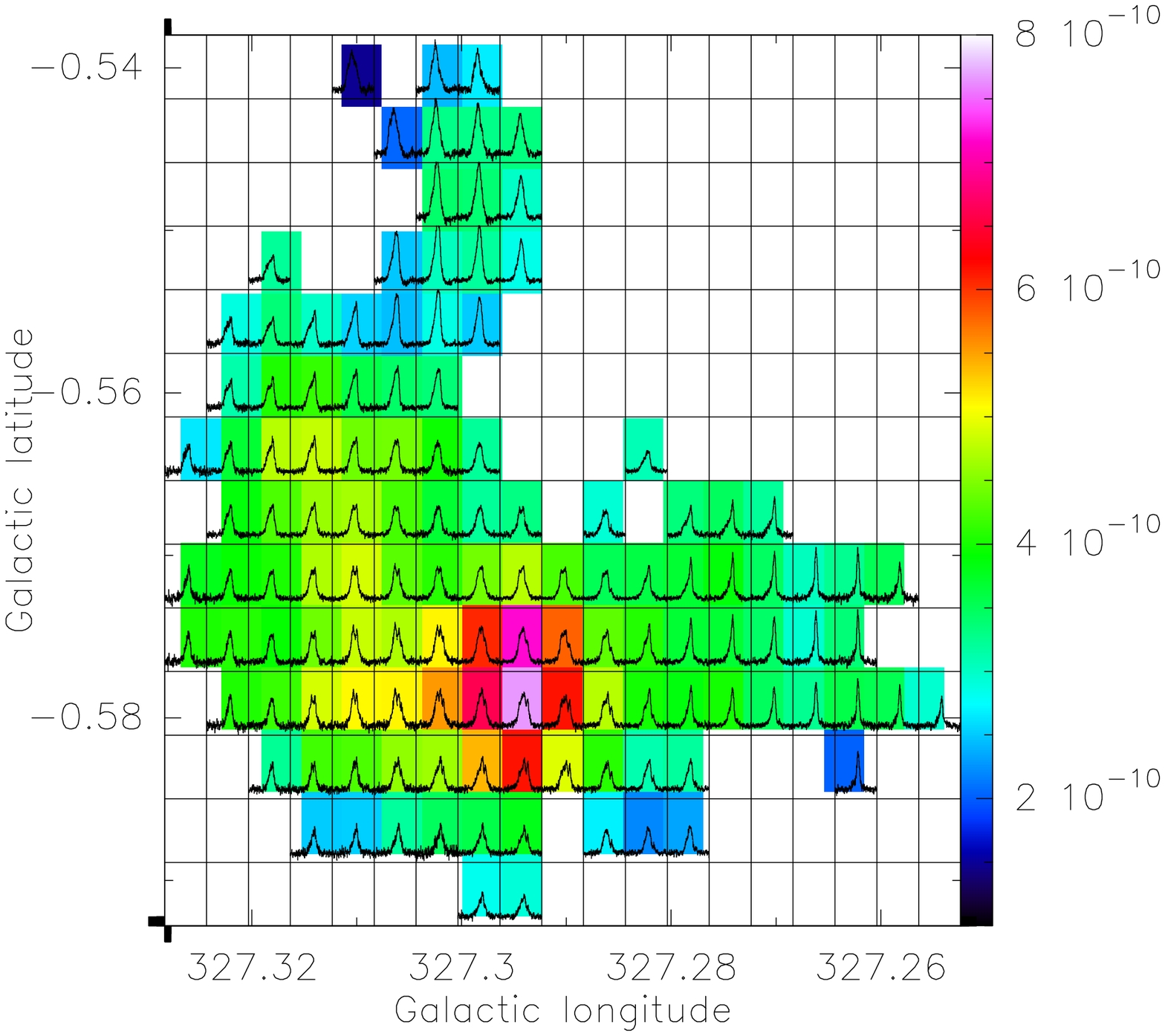,width=3.5in,height=3.3in}
\psfig{file=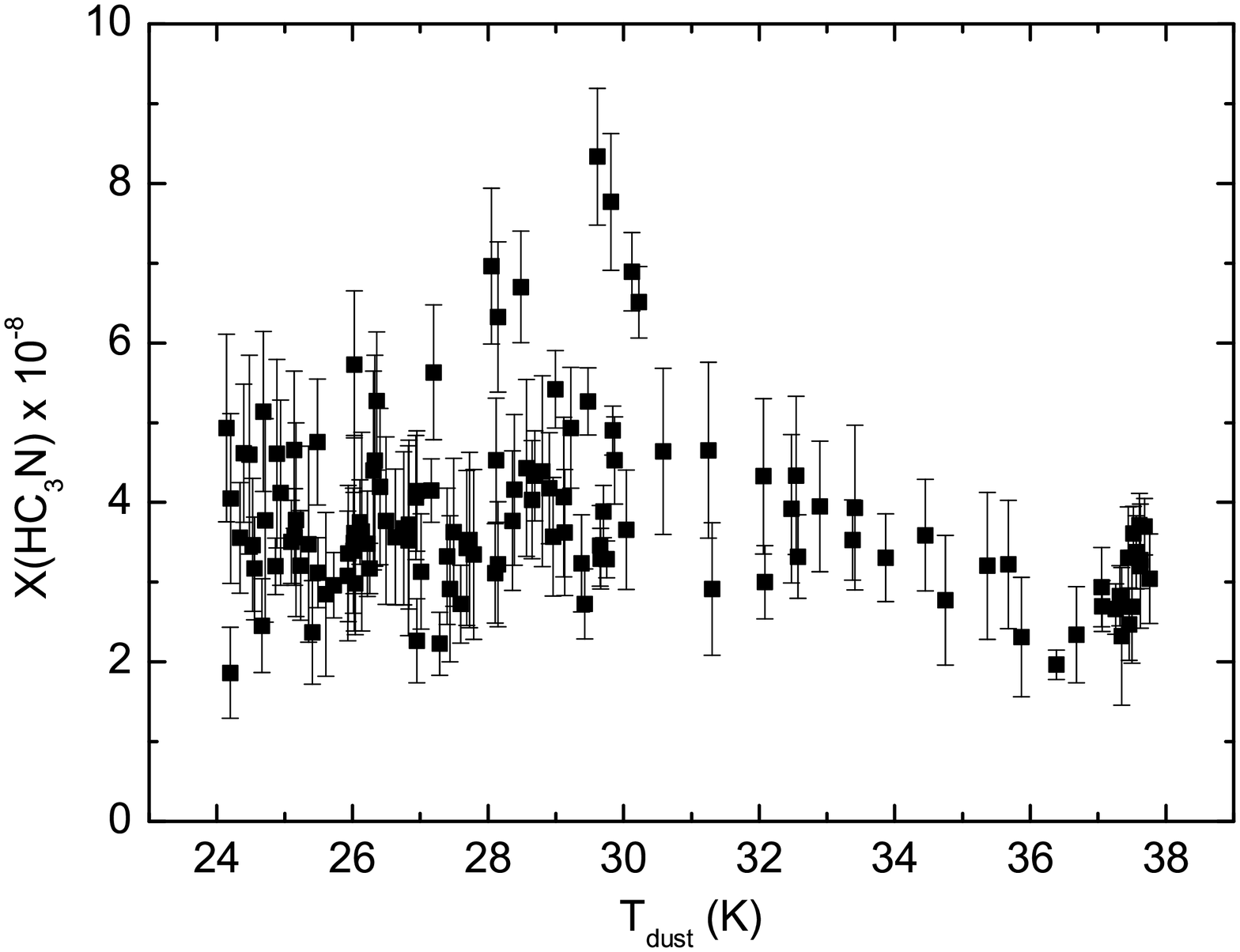,width=4.5in,height=3.3in} \caption{Top:
The HCO$^+$ (1-0) spectra superimposed on the calculated HC$_3$N
abundance map of G327.293-0.579. Bottom: Abundance of HC$_3$N
plotted as a function of the dust temperature in each pixel of
G327.293-0.579. }
\end{figure*}

\begin{figure*}
\centering
\psfig{file=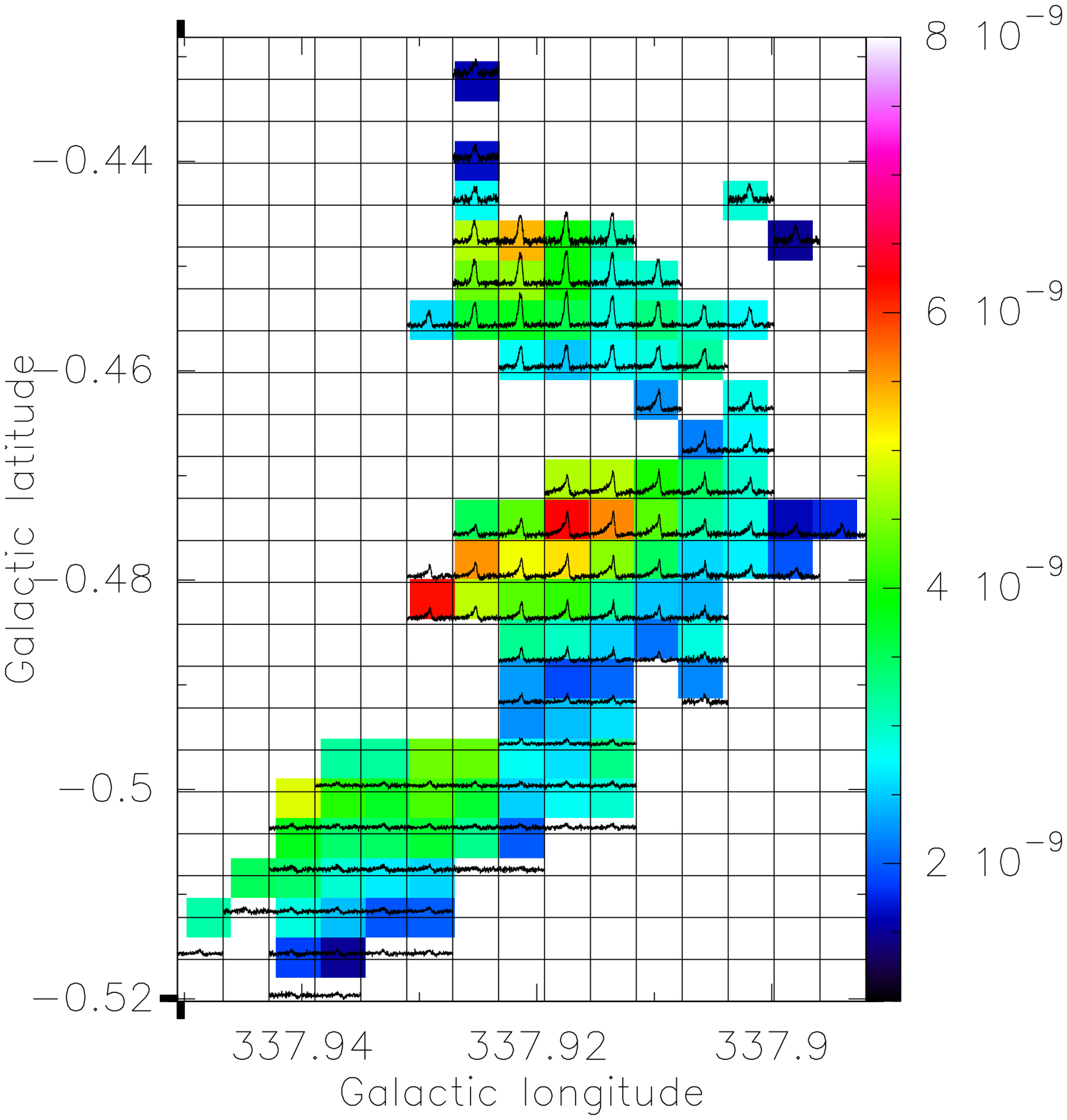,width=3.5in,height=3.3in}
\psfig{file=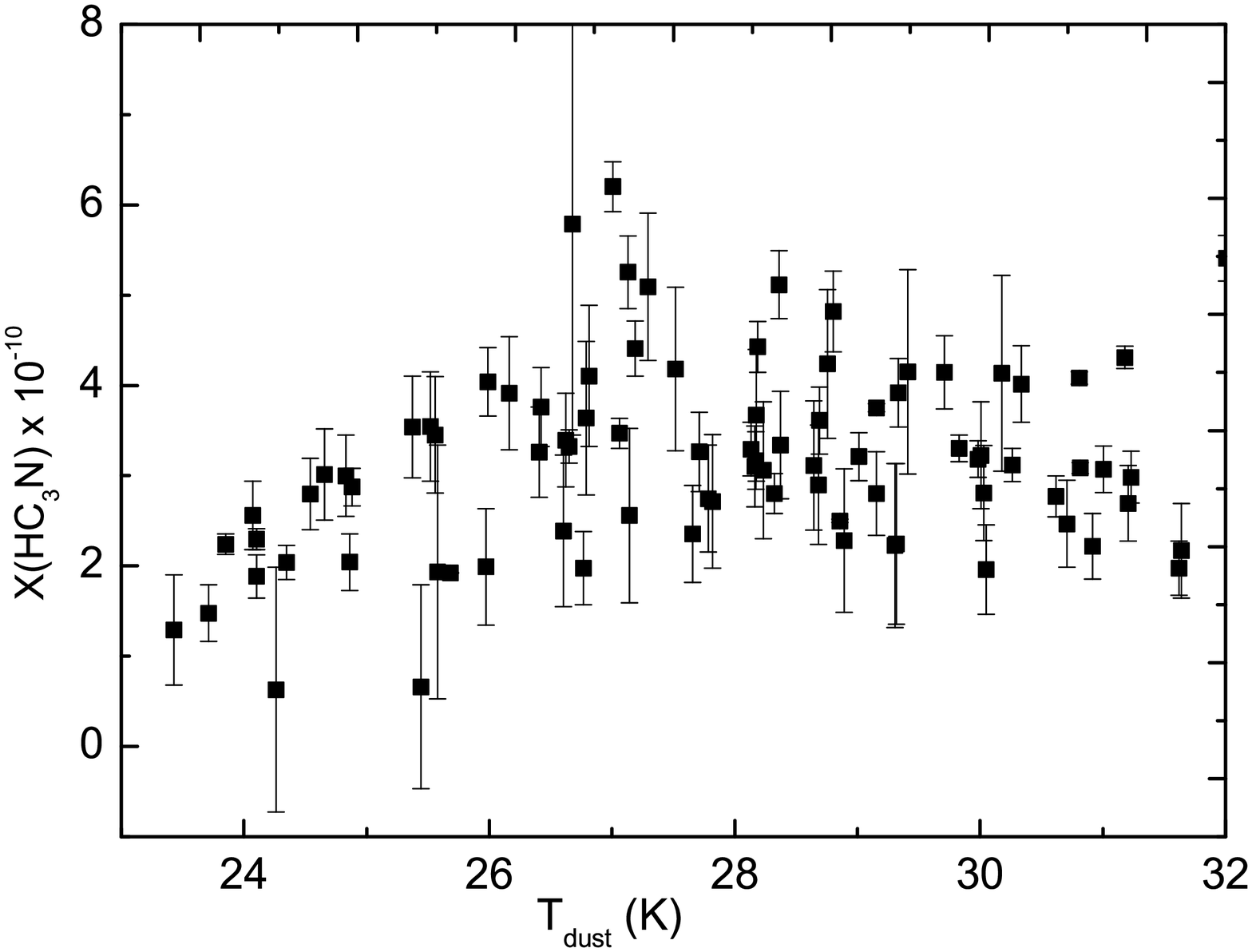,width=4.5in,height=3.3in} \caption{Top:
The HCO$^+$ (1-0) spectra superimposed on the calculated HC$_3$N
abundance map of G337.916-0.477. Bottom: Abundance of HC$_3$N
plotted as a function of the dust temperature in each pixel of
G337.916-0.477. }
\end{figure*}

\clearpage
\begin{figure*}
\centering
\psfig{file=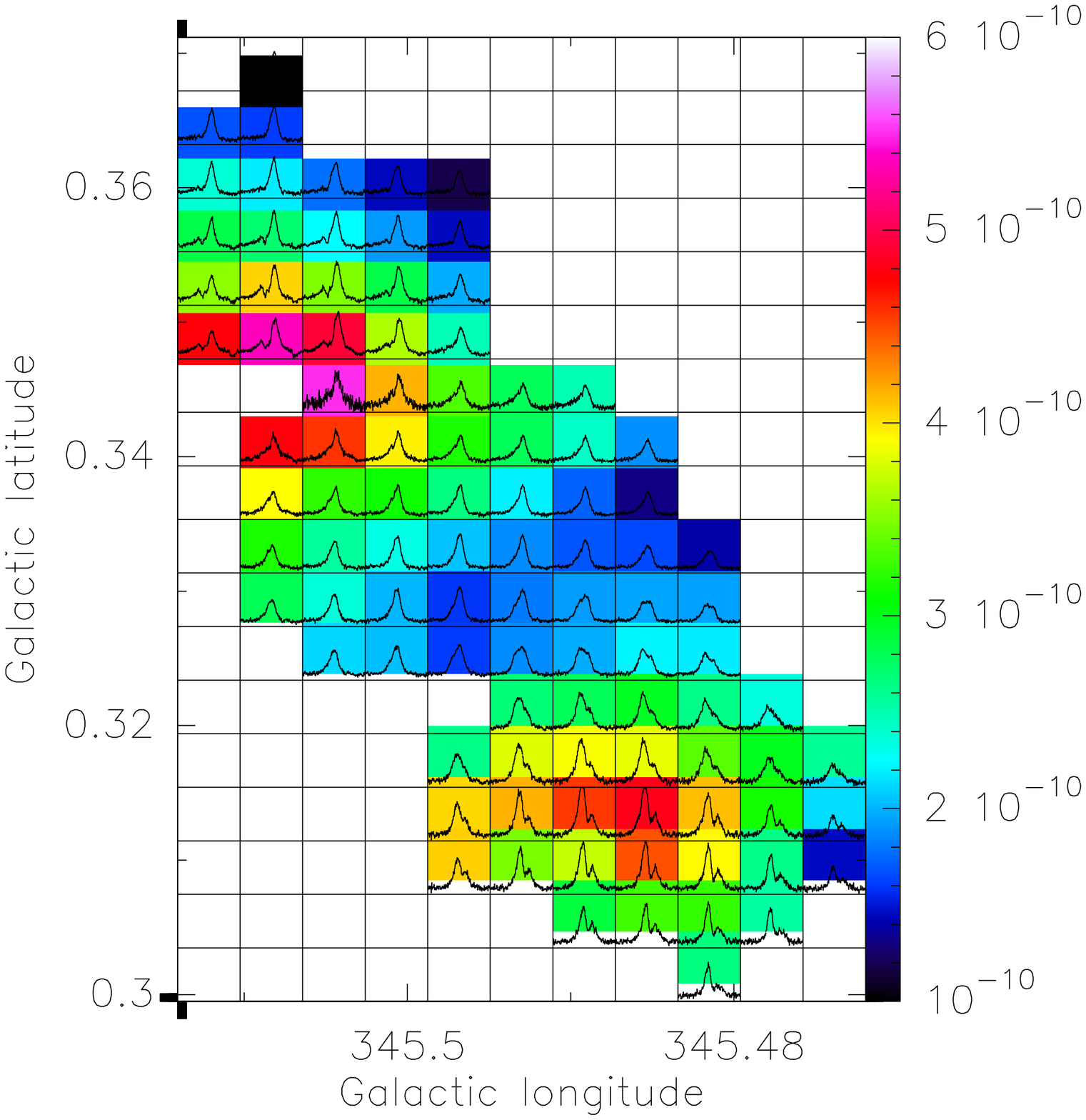,width=3.5in,height=3.3in}
\psfig{file=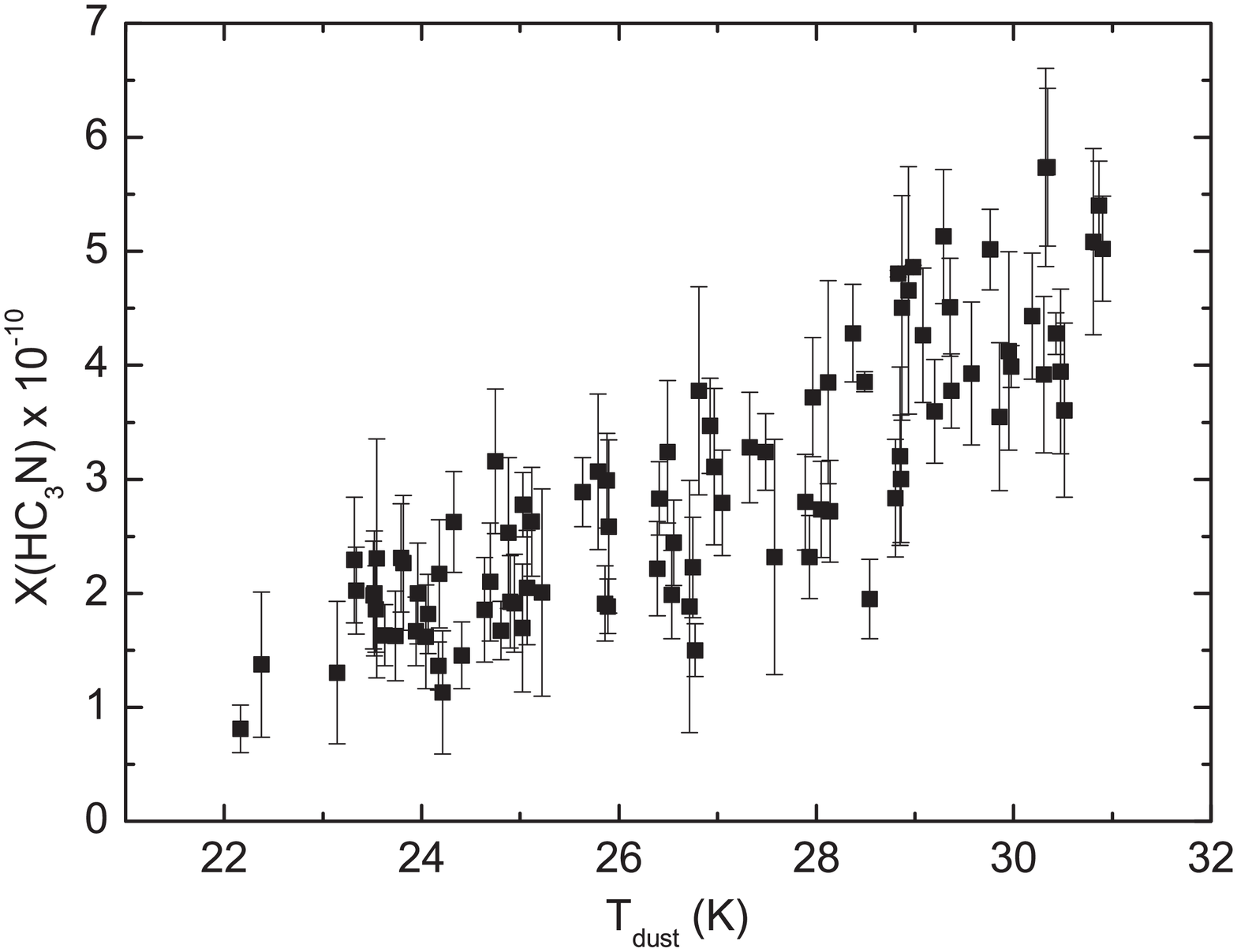,width=4.5in,height=3.3in} \caption{Top:
The HCO$^+$ (1-0) spectra superimposed on the calculated HC$_3$N
abundance map of G345.448+0.314. Bottom: Abundance of HC$_3$N
plotted as a function of the dust temperature in each pixel of
G345.448+0.314. }
\end{figure*}

\begin{figure*}
\centering
\psfig{file=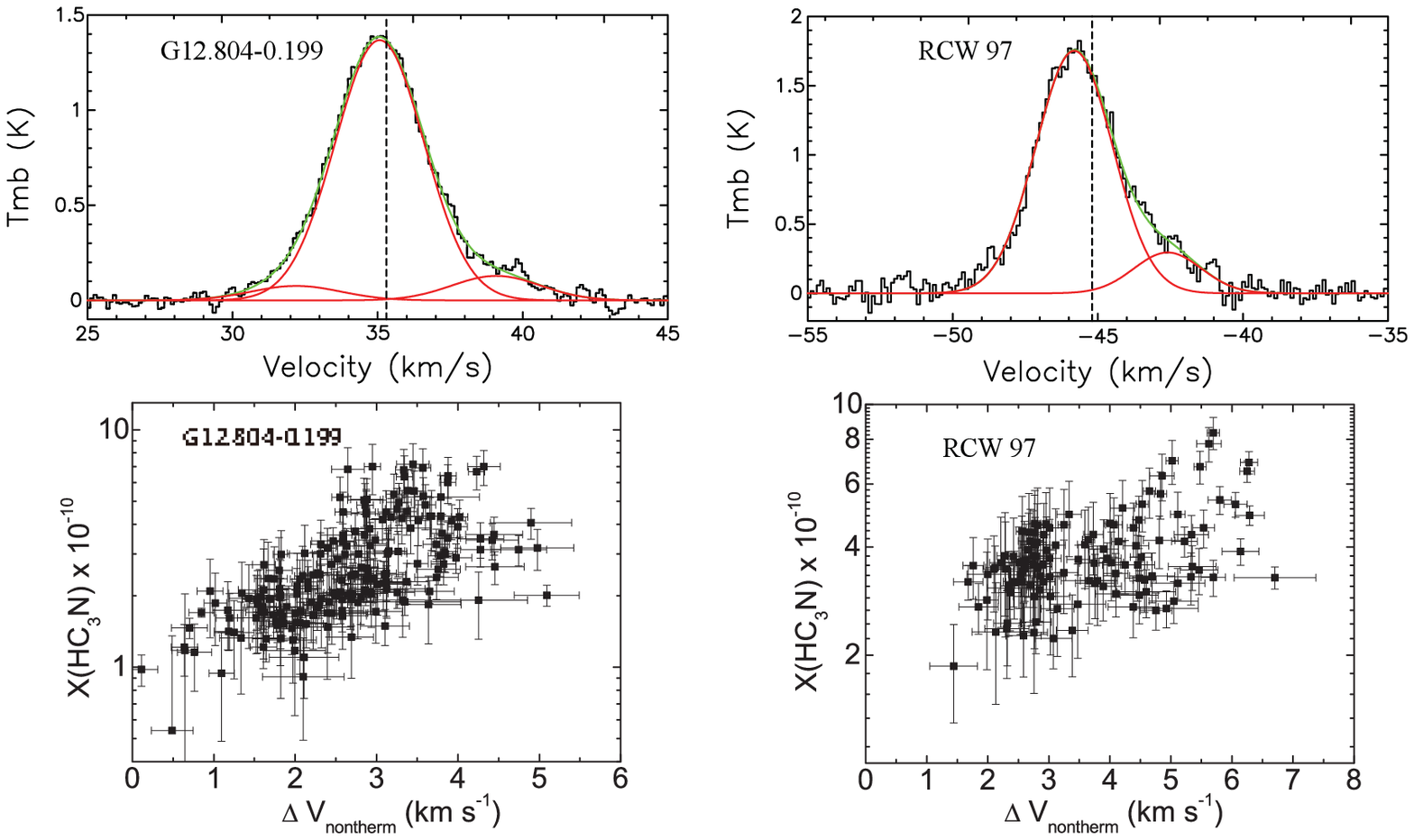,width=6in,height=7in} \caption{Top panels: The averaged spectra of HC$_3$N (10-9)
in G12.804-0.199 and RCW 97 where its abundance is highest. Gaussian fits
to individual components are plotted using solid red lines and the cumulative fits with solid green lines. Vertical
dashed lines mark the systemic velocity of each source measured from the Gaussian fit to their N$_2$H$^+$ (1-0) lines. Bottom panels:
The abundance of HC$_3$N plotted as a function of its nonthermal line
width in G12.804-0.199 and RCW 97.}
\end{figure*}

\begin{figure*}
\centering
\psfig{file=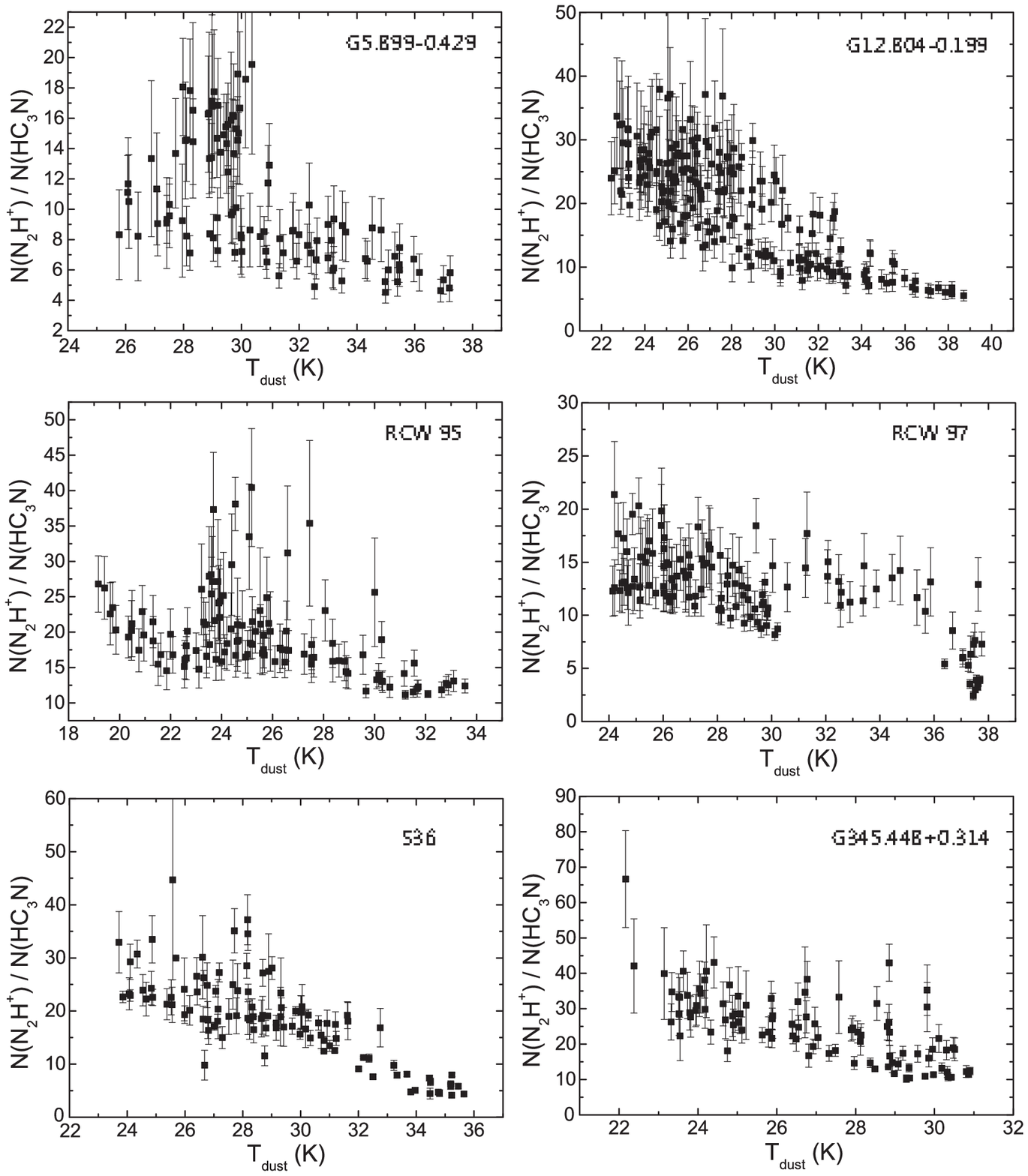,width=6in,height=8in} \caption{The
relationship between $N(N_2H^+)$/$N(HC_3N)$ and $T_{dust}$ in all of our sources.
For RCW 95 and S36, the column densities of N$_2$H$^+$ are from YXW18. For other sources,
we calculated their column densities of N$_2$H$^+$ in the same way.}
\end{figure*}

\end{document}